\documentclass[11pt,a4paper]{article}

\usepackage[utf8]{inputenc}         
\usepackage[T1]{fontenc}            
\usepackage{lmodern}                
\usepackage{setspace}               
\usepackage[margin=1in]{geometry}   
\usepackage{graphicx}               
\usepackage{amsmath,amssymb,amsfonts,amsthm} 
\usepackage{bm}                     
\usepackage{mathtools}              
\usepackage{microtype}              
\usepackage{booktabs}               
\usepackage{caption}                
\usepackage{subcaption}             
\usepackage{enumerate}              
\usepackage{xcolor}                 
\usepackage{hyperref}               
\usepackage{natbib}
\usepackage{url}                    
\usepackage{authblk}                
\usepackage[markup=underlined]{changes} 
\usepackage{todonotes}
\usepackage{ulem}
\usepackage{placeins}
\hypersetup{
    colorlinks = true,
    linkcolor  = blue,
    citecolor  = blue,
    urlcolor   = blue,
}

\title{Planning difficulty and temporal constraints differently shape the level of detail and ordering of visual exploration}

\author[1  ]{Mattia Eluchans}
\author[1,*]{Giovanni Pezzulo}

\affil[1]{Institute of Cognitive Sciences and Technologies, National Research Council, Rome, Italy}
\affil[*]{Corresponding author: giovanni.pezzulo@istc.cnr.it}

\begin{document}
\maketitle

\begin{abstract}

Planning entails identifying sequences of actions to reach a goal, but how much of a future plan is constructed before movement and how visual exploration supports this process remain poorly understood. Here, we investigated how visual exploration before and during action relates to the executed path and how this relationship is shaped by task difficulty and temporal constraints. Participants solved a path-finding task while their gaze and navigation movements were recorded. We manipulated planning difficulty using \textit{misleadingness}, which captures the mismatch between the perceptual layout of a problem and the structure of its solution, and manipulated temporal constraints by either allowing immediate movement or imposing a temporal delay.
We quantified how visual exploration covered the executed path (spatial coverage) and retained its sequential organization (temporal coherence). Greater difficulty increased visual exploration, repeated sampling, and spatial coverage, but was associated with lower temporal coherence with the upcoming path. When participants could act immediately, pre-movement visual exploration showed lower spatial coverage than during execution. Imposing a delay increased pre-movement spatial coverage, bringing it closer to the level observed during execution, but did not increase temporal coherence. The delay was also associated with fewer pauses and backtracks, suggesting that plans were better consolidated before movement, reducing the need for online revision.
Individual differences reflected continuous variation in the balance between pre-movement planning and online revision. Together, these findings suggest that participants flexibly allocate visual exploration before and during plan execution, adjusting its level of detail and ordering to problem difficulty and temporal constraints.

\end{abstract}
\textbf{Keywords:} eye-tracking; planning; adaptive behaviour; temporal ordering; visual exploration

\newpage

\section*{Introduction}

Planning can be defined as the cognitive process of generating a sequence of steps that leads to a specific goal \citep{newell1972human, hayes1979cognitive, mattar2022planning, sacerdoti1974planning,donnarumma2016problem,maisto2015divide}. While human planning has been studied since the origins of cognitive science, the process leading to the construction of plans -- and the temporal sequencing of plan elements -- is covert and hence remains difficult to access experimentally. 

As a consequence of this limitation, cognitive and behavioural studies often rely on indirect measures of planning processes. An influential approach, for example, uses reaction times as a proxy for planning time, linking them to the amount of information that must be processed \citep{stone1960models, ratcliff1978theory, Fudenberg2019Testing, Myers2022A, simonelli2025structuring, simonelli2026foraging}, or to the depth of the plan required to solve the task \citep{Proctor2018}. Other approaches focus on characterizing the observable plans implemented by participants through model-based analyses \citep{Campitelli2004, lancia2023humans, chen2026justtimeworldmodeling, kessler2024human, ho2022people,  rens2023entropy}, and typically compare participants’ choices -- either globally or step-by-step -- with those of normative models and artificial agents. The increasing capabilities of artificial agents have further expanded the range of tasks for such comparisons \citep{Allen2024GamesMind, FernandezVelasco2024, vanOpheusden2023Expertise}. 

A complementary perspective focuses on how choice and planning dynamics unfold in time, by tracking participants' movement (or cursor) kinematics during decision-making and problem solving \citep{barca2012unfolding, barca2015tracking, spivey2008continuity, spivey2005continuous, buc2017continuous}. In parallel, research in motor control identifies anticipation and coarticulation -- changes to an action based on upcoming ones -- as key markers of sequential planning in tasks that are routinized, instructed, or planned from scratch \citep{yoo2021continuous, Reina2003, Eluchans2025EyeHand,maselli2025whole}.

Gaze behaviour can provide an even more direct proxy for sequential decision and planning processes \citep{LAKSHMINARASIMHAN2020662}. Studies have reported that while low-level visual salience attracts fixations to regions of high contrast or conspicuity \citep{mannan1997fixation, itti2001computational, parkhurst2002modeling, parkhurst2003scene}, task demands consistently override purely stimulus-driven guidance. For example, gaze patterns tightly align with behavioural goals across naturalistic contexts, enabling accurate inference of ongoing tasks and revealing structured, purposeful sampling of the environment \citep{rothkopf2016task, hayhoe2011vision, hayhoe2005eye}. Anticipatory gaze also reflects developing internal knowledge: as people learn statistical or rule‑based structure in visual tasks, their fixation patterns gradually reorganize to mirror that structure, indicating that eye movements participate in active learning itself \citep{arato2024eye}. Relatedly, in insight problems, gaze reveals key intermediate states of reasoning, predicting who will discover solutions and facilitating problem restructuring when attention is externally guided \citep{grant2003eye}. 

Several studies have also reported that gaze actively supports upcoming behaviour across diverse tasks by implementing ‘look-ahead’ strategies to anticipate future actions. For example, people pre‑allocate attention along planned (manual) trajectories \citep{Baldauf2018}, and even distribute attention across multiple upcoming movement targets before any action begins, indicating parallel planning of sequential movements \citep{BALDAUF20064355, BALDAUF2010999}. During everyday actions, such as handwashing, a consistent portion of fixations is directed towards objects relevant only for future steps \citep{PELZ20013587}. During the observation of grasping actions, people make anticipatory saccades to the object they infer the other person will grasp \citep{ambrosini2011grasping,donnarumma2017action}. In sports, skilled performers exhibit predictive eye movements tailored to the structure of the task \citep{land2000eye}; furthermore, gaze predicts physical movements, such as in ballistic interception tasks \citep{Gerstenberg2017, Huber2004}. In linguistic and musical tasks, reading gaze leverages a short‑term memory buffer to anticipate action demands seconds before they occur \citep{land1997knowledge,donnarumma2025integrating}. 

Recent work has addressed the potential roles of these ‘look-ahead’ strategies in supporting sequential planning and problem solving, revealing a range of patterns. During pre‑navigation planning, gaze exhibits a broad-to-narrow search pattern, reduced saccade amplitudes, and increased fixations as subgoals are identified, with attention concentrated on map locations most critical for connectivity \citep{gordon2026gaze}. These gaze signatures track the emergence of hierarchical decomposition and predict planning success in complex environments. During multi‑step physical problem solving, observers choose visual subgoals that minimize present and future cognitive costs, revealing a sophisticated integration of immediate demands with anticipated difficulty \citep{binder2025humans}. 
In another problem-solving task, gaze position was predictive both of the future cursor movements and of the next fixation, suggesting that participants planned multiple eye movements in advance \citep{Eluchans2025EyeHand}. Additional evidence that consecutive fixations are not executed independently comes from model-based analyses of visual search \citep{Hoppe2019} and normative accounts of active sensing, in which gaze samples the environment in order to reduce uncertainty \citep{najemnik2005optimal, Yang2016}. Related work has characterized gaze behaviour as reflecting a trade-off between the expected benefits of accurate perception and the costs of information foraging \citep{Wagner2023}. Evidence for look-ahead gaze strategies has also emerged from navigation tasks in which the exit location was either unique or known in advance to the observer \citep{ZhuElifeEYEMovementPlanning, li2023modeling}.

Despite this extensive body of work, we still lack a complete understanding of how visual exploration before movement supports plan formation and determines its level of detail. A key question is the extent to which visual exploration prior to movement aligns with -- and thus anticipates -- the subsequently executed plan (e.g., the path taken during spatial problem solving). Do participants visually explore the entire path they will later execute? Does the order of fixations correspond to the order of the subsequently executed path? A further question is whether the level of detail at which plans are formed reflects a fixed strategy or is flexibly modulated by problem difficulty, temporal constraints, and individual differences.

This work addresses these open questions by analysing participants’ eye and navigation movements during the ThinkAhead problem-solving task \citep{Eluchans2025AdaptivePlanning,Eluchans2025EyeHand}. In this two-dimensional, graph-based path-finding task, participants must trace a route that, starting from a given point, visits a set of targets without passing through the same node twice, within a limited time (60 seconds). Given this revisit inhibition, participants cannot simply move as fast as possible to reach all the targets, as a greedy strategy would likely lead them to a dead end. Successful performance, therefore, requires some degree of planning.

Crucially, we manipulate both problem difficulty and temporal constraints. To manipulate problem difficulty, we introduce and validate a new task property -- \textit{misleadingness} -- that captures the cumulative mismatch between the perceptual proximity of nodes and their relative graph-theoretic positions along the solution path. Under this measure, problems are easy (difficult) when nodes that are close in physical space are also close (distant) along the solution path. To manipulate temporal constraints, we present problems in two counterbalanced blocks. In one block, participants can begin moving as soon as the problem appears on the screen; in the other, we impose a 19-second delay before movement can begin. These manipulations allow us to assess the level of detail in visual exploration before movement and the extent to which it aligns with the subsequently executed plan. We evaluate this alignment along two complementary dimensions: \emph{spatial coverage}, measuring how much of the future path was fixated, and \emph{temporal coherence}, measuring how closely the order of fixations matched the order of the subsequently executed path. We then examine how these measures vary as a function of task difficulty and the presence or absence of a delay period.

We hypothesize that, if planning is resource-rational \citep{Callaway2022}, relying on simplified representations \citep{ho2022people} and limited planning horizons \citep{kryven2024approximate}, visual exploration before movement will not fully cover and temporally align with the subsequently executed plan. However, because planning time and depth can flexibly adapt to problem difficulty \citep{Eluchans2025AdaptivePlanning}, we expect greater difficulty to elicit more extensive pre-movement planning, reflected in longer inspection times, greater visual exploration of alternative routes, and more frequent refixations. This increased investment could produce a more detailed and temporally coherent gaze-based representation of the subsequently executed path, reflected in greater \emph{spatial coverage} and \emph{temporal coherence}.

We further hypothesize that temporal constraints will modulate the level of pre-movement planning. While people often begin acting with incomplete plans that are refined online \citep{nuzzi2026planning, lepora2015embodied, maselli2023beyond}, imposing a delay before movement should allow more planning to be completed before movement onset, resulting in greater \emph{spatial coverage} and \emph{temporal coherence}, whereas immediate movement should promote partial pre-movement planning followed by greater reliance on online correction. Finally, given substantial individual differences in planning \citep{lancia2023humans,krichmar2023importance} and eye movements \citep{zangrossi2021visual}, we expect individuals to differ in spatial coverage and temporal coherence, as well as in the extent to which they rely on pre-movement planning versus online correction.

\section*{Methods}
\label{sec:methods}
\subsection*{Participants} For the experiment we recruited 28 participants (12 F, 16 M; $mean = 29 \pm 5 \ y$, range = $[19,42] \ y $) with normal or corrected-to-normal vision.  

\subsection*{Ethics statement} All participants gave informed consent to our procedures which were approved by the Ethics Committee of the National Research Council. Participants were free to leave the experiment at any moment. 

\subsection*{The task} 

\begin{figure}[ht!]
    \centering
    \includegraphics[width=1\linewidth]{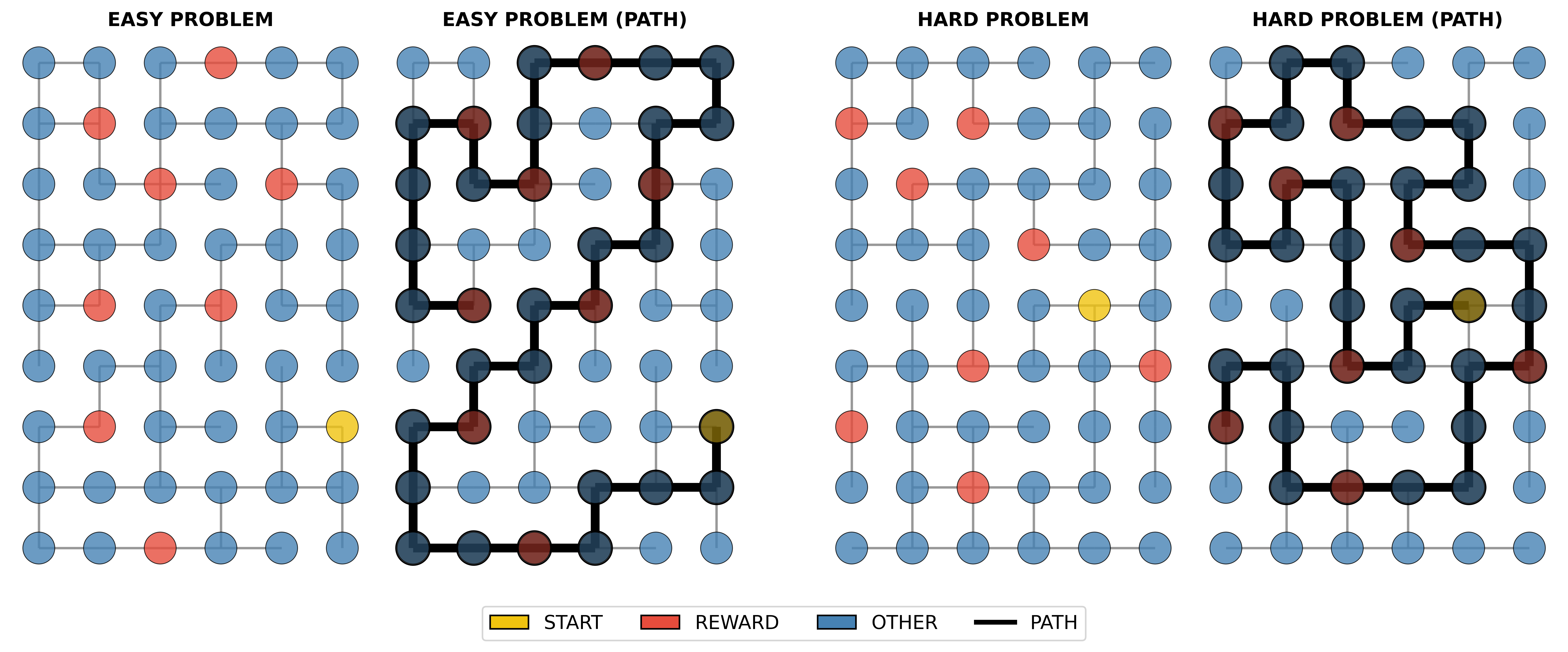}
    \caption{Comparison of representative problem instances having low misleadingness (\emph{easy} problems) and high misleadingness (\emph{hard} problems). Left to right: easy problem, easy problem with highlighted path, hard problem, and hard problem with highlighted path. Nodes indicate start (yellow), target (red), and other nodes (steelblue); the semi-transparent black overlay marks the selected path.}
    \label{fig:easy_hard_path_comparison}
\end{figure}

Participants completed 60 problems, each requiring them to find a path starting from a given (yellow) node and connecting all $8$ (red) coloured “rewards” on a (not-fully connected) grid, made by 54 nodes arranged in a $6\times9$ structure (see \autoref{fig:easy_hard_path_comparison} for example problems). Passing twice over the same node was inhibited, but it was possible to backtrack and unselect nodes by tracing back from the last position. Each problem had to be solved within 60 seconds. The time countdown was shown at the top-right of the screen, as well as the number of problems remaining to reach the next block, the number of problems unsolved and the total score.

The 60 problems were organized in two interleaved blocks, comprising 30 randomized problems each with the constraint that half were easy and half were hard in each block. The specific easy and hard problems assigned to the first and second blocks were randomized. The time after which movement was permitted changed between blocks: in one block, participants could start moving as soon as they wished; in the other block, they had to wait for 19 seconds before starting (a 4-second countdown, with a clear sound played each second, began after 15 seconds). The order of conditions was counterbalanced across participants. This led to a $2\times2$ within-subjects design, with block (no-delay vs. delay) and problem difficulty (easy vs. hard) as experimental factors.

\subsection*{Problem selection and definition \emph{misleadingness} as a measure of difficulty}

We selected a set of problems with the same number of nodes and rewards, and whose shortest solutions' properties (solution length, number of turns, number of points where a deviation made the task unsolvable) took values in a very narrow set (\autoref{fig:struct_prop}). The key dimension we manipulated was task difficulty, defined here by introducing what we called \textit{misleadingness}: a novel path property defined as the cumulative mismatch between the graph-distance (number of links) between two nodes of a solution and the distance between their indexes within the solution. 

Formally, consider a map represented as an undirected graph $G = (V, E)$, and let
$\mathcal P = (p_1, p_2, \dots, p_n)$ denote the shortest solution path (i.e., the
shortest sequence of nodes visiting all reward nodes from the start node).
For each node $p_i \in \mathcal P$, we identify its \emph{forward neighbors}: the set
of graph neighbors that also belong to the solution and appear later in the
sequence,
\begin{equation}
    \mathcal{N}^{+}(p_i) = \bigl\{ p_j \in \mathcal P \,:\, (p_i, p_j) \in E
    \;\text{and}\; j > i \bigr\}.
\end{equation}
For each such forward neighbor $p_j$, the \emph{temporal distance}
$|j - i|$ measures how far apart the two nodes are in the solution ordering.
A large temporal distance indicates that the neighbor, while spatially
adjacent, is reached much later in the optimal path---making it a potentially
misleading choice at that decision point.

The \emph{node-level misleadingness} of $p_i$ is defined as the sum of temporal distances to all its forward neighbors:
\begin{equation}
    m(p_i) = \sum_{s_j \in \mathcal{N}^{+}(p_i)} |j - i|.
\end{equation}
The \textit{Misleadingness} is then obtained by summing over all nodes in the shortest solution:
\begin{equation}
    M(\mathcal P) = \sum_{i=1}^{n} m(p_i) = \sum_{i=1}^{n} \sum_{p_j \in
    \mathcal{N}^{+}(p_i)} |j - i|.
\end{equation}

For this exploratory study we considered two ranges of values of this property (less than $10$ and more than $100$), to ensure a marked difference in difficulty output:
\begin{equation}
    \text{Difficulty} =
    \begin{cases}
        \textit{easy} & \text{if } M(\mathcal P) < 10, \\
        \textit{hard} & \text{if } M(\mathcal P) > 100.
    \end{cases}
\end{equation}

Intuitively, a high $M(\mathcal P)$ indicates that the map contains many decision points where spatially adjacent nodes lead to distant segments of the optimal path. A navigator following a locally greedy strategy (visiting the nearest unvisited neighbor) would be frequently ``misled'' into suboptimal detours, making planning ahead essential for efficient navigation. 
Selected problems were also chosen such that the shortest solution was also the one having the minimum value of misleadingness. Examples of problems with low misleadingness (henceforth, \emph{easy}) and high misleadingness (henceforth, \emph{hard}), along with their solutions, are shown in \autoref{fig:easy_hard_path_comparison}.

\subsection*{Experimental setup} 
To track eye movements, we used an EyeLink 1000+ tracker (35 mm lens) in a chin-forehead rest desktop-mount configuration, set according to the manufacturer’s instructions: the chin-forehead support was kept at a fixed height such that the eyes of the subjects were approximately at $\frac{3}{4}$ of the screen height. Subjects were seated on a height-adjustable chair. The experiment was performed inside an isolated room with no sources of light other than the IR-camera of the gaze tracker and the screen where the task app was running. The IR-camera was placed between the subject and the screen, at approximately 50 cm from participants’ eye position.

\subsection*{Experimental protocol}  Participants were first shown a video tutorial and given information about the structure of the experiment: (1) calibration; (2) validation; (3) 4 training problems; (4) 60 problems of the task. Participants were instructed that their total score depended on problem completion and the remaining time: the faster the completion, the greater the number of points they earned. An unsolved problem returned 0 points. The passage from the training block to the first block, as well as the transitions from the first block to the second, and the end of the whole experiment, were indicated by a message on the screen.
For each participant, a standard 9+1-point calibration for eye tracking was performed  at the beginning of the experiment on a grid  overlapping with the area of the task problems. A 9+1-point validation was then performed for the eye tracking calibration, after which participants were asked to move as little as possible.

\subsection*{Exclusion criteria} Trials in which participants began moving the cursor before the end of the countdown were excluded from the analysis. One participant (P19) was excluded because of systematic violations of the countdown condition (27 trials). Among the remaining participants, 1555 of the 1620 recorded trials were retained for analysis.

\subsection*{Mouse tracker data preprocessing}

\begin{figure}
    \centering
    \includegraphics[width=1\linewidth]{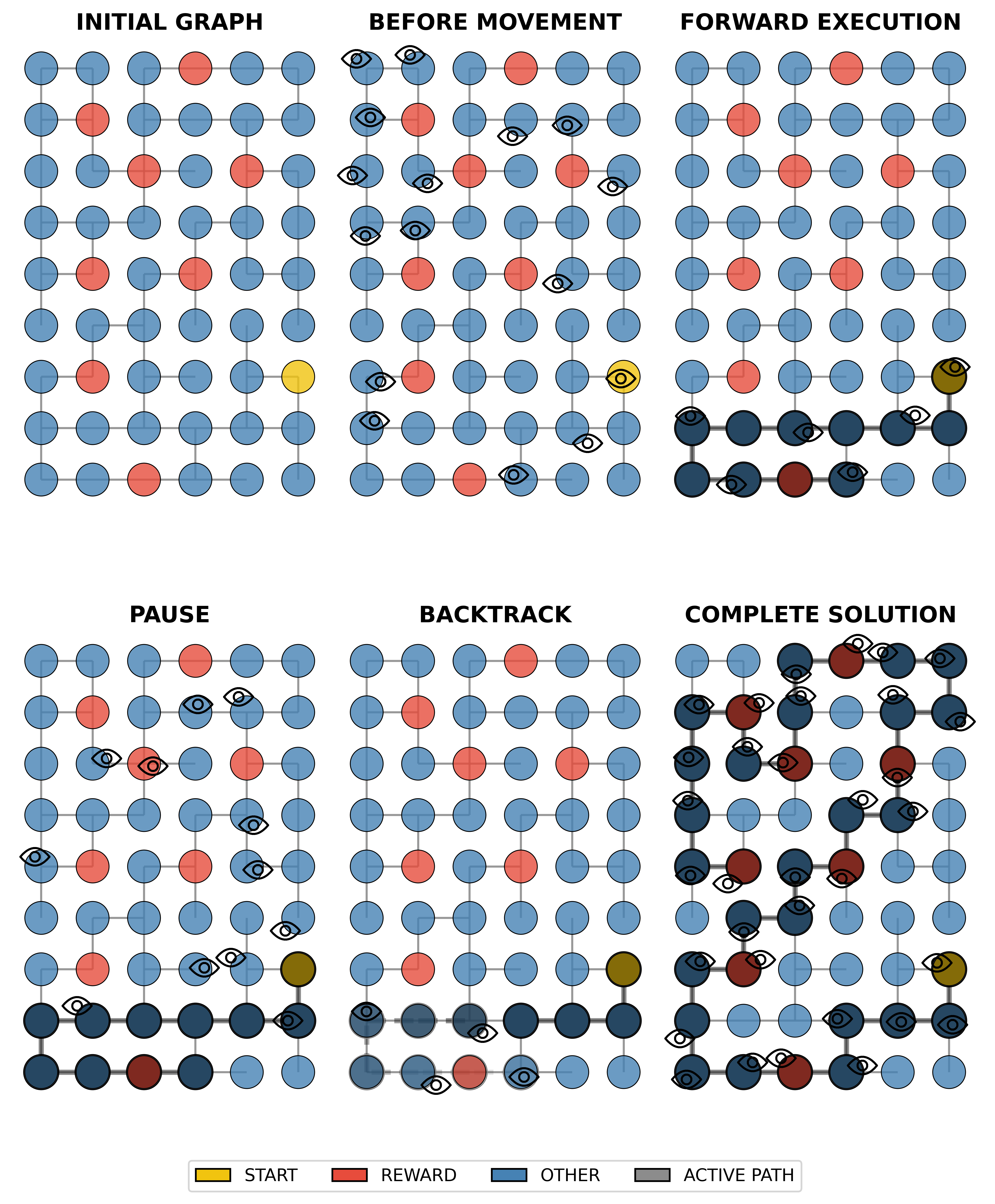}
    \caption{Representation of a typical task progression. At first the problem is presented on the screen (Initial graph), and there is a phase where participants visually explore the map, ideally looking for a solution (Before movement).  A partial path is implemented (Forward execution), until a mistake is recognized and the implementation is stopped, replaced by another visual exploration (Pause). The error is corrected by tracing back some steps (Backtrack; unselected nodes are partially shaded) and, finally, the correct solution is implemented (Complete solution). Eyes represent fixation locations, whereas the colored nodes represent start (yellow), reward (red), and other (azure) nodes. Darker nodes indicate the selected path.}
    \label{fig:progression_of_a_trial}
\end{figure}

The cursor position was recorded throughout the experiment at a sampling frequency of 60 Hz. These trajectories were used to reconstruct velocity profiles which, as reported in previous work \citep{Eluchans2025EyeHand}, exhibit a bimodal distribution: a first peak corresponding to very low velocity values and a second peak associated with movement execution. Because this distribution varies across participants, we determined a participant-specific threshold separating the two peaks, defined as the first local minimum in the velocity distribution between them.
Pauses were defined as temporal intervals during which mouse velocity remained below this threshold for at least 100 ms. Based on this criterion, four possible events were identified within each trial: the pre-movement phase, pauses, the execution phase, and the backtrack phase (\autoref{fig:progression_of_a_trial}). In the present paper, results related to the backtrack phase were not considered.

\subsection*{Eye tracker data preprocessing}
The acceptance criteria for the calibration were those suggested by the manufacturer: a maximum error in calibration of 1° and an average error smaller than 0.5°. The eye tracking sampling frequency was 2000 Hz.
Across participants, the mean calibration errors were $\Delta(L) = 0.300^\circ \pm 0.019^\circ~ (SE;\ range : [0.14^\circ, 0.49^\circ])$, for the left eye and $\Delta(R) = 0.295^\circ \pm 0.016^\circ~(SE; \ range : [0.16^\circ, 0.54^\circ])$, for the right eye. The maximum errors, averaged across participants, were $\Delta_{\max}(L) = 0.576^\circ \pm 0.034^\circ~(SE; \ range : [0.23^\circ, 0.90^\circ])$, for the left eye and $\Delta_{\max}(R) = 0.568^\circ \pm 0.026^\circ~(SE;\ range : [0.32^\circ, 0.95^\circ])$, for the right eye.
All gaze data were preprocessed before analysis to convert the raw eye‑tracking signal into time‑ordered sequences of fixations aligned with the task graph. Fixations and saccades were identified using the EyeLink 1000+ online parser with the standard \textit{cognitive configuration}: a velocity threshold of $30^\circ/s$, an acceleration threshold of $8000^\circ/s^2$, and a motion threshold of $0.15^\circ$.

For each fixation, gaze coordinates were mapped onto the nearest graph node, with off‑graph fixations (e.g., towards the timer or score displays) labeled separately. Consecutive fixations assigned to the same node were merged.
Across participants, the mean number of fixations per trial was $72.0 \pm 19.8$ ($\pm$ SD; range: $41.9$--$135.9$) for easy problems without  delay, $98.8 \pm 17.0$ (range: $70.6$--$145.0$) for easy problems with  delay, $123.8 \pm 22.9$ (range: $80.0$--$177.3$) for hard problems without  delay, and $136.3 \pm 24.7$ (range: $91.0$--$175.6$) for hard problems with  delay.

\subsection*{Characterization of performance and visual exploration}
In each trial, participants’ behaviour was characterized along three complementary dimensions. First, we assessed task performance and the temporal distribution of behaviour, including whether the problem was successfully solved, the time spent inspecting it before movement onset, and the numbers of pauses and backtracks produced during execution. These variables could be obtained directly from the preprocessed data. Second, we characterized visual exploration across task phases (i.e., before movement onset, during execution and during pauses) independently of the implemented solution, considering the number and distribution of fixations across node types and the frequency with which participants revisited previously inspected nodes (\emph{re-fixation proportion}). Finally, we examined the alignment between visual exploration and the executed path, quantifying both how much of the upcoming path was fixated (\emph{spatial coverage}) and how closely fixation order matched execution order (\emph{temporal coherence}). These novel measures (\emph{re-fixation proportion}, \emph{spatial coverage} and \emph{temporal coherence}) are described in the following sections.

\subsection*{Re-fixation proportion (RP)}

To quantify the extent to which participants' gaze revisited previously inspected locations, we computed a \emph{Re-fixation Proportion (RP)} for each trial phase. Let $N_{\text{fixations}}$ denote the total number of fixations in the phase, and $N_{\text{unique}}$ the number of distinct graph nodes that were fixated at least once. The re-fixation proportion is then defined as:

\begin{equation}
RP = 1 - \frac{N_{\text{unique}}}{N_{\text{fixations}}}.    
\end{equation}

Values range from $0$ to $1$, where $RP = 0$ indicates that all fixations fell on different nodes (maximal diversity), and values approaching $1$ indicate a high prevalence of re-fixations on previously viewed nodes. This measure provides a simple, phase-specific index of how much gaze behaviour involves returning to the same elements of the map.

\subsection*{Spatial coverage and temporal coherence between visual exploration and the executed path}

\begin{figure}
    \centering
    \includegraphics[width=1\linewidth]{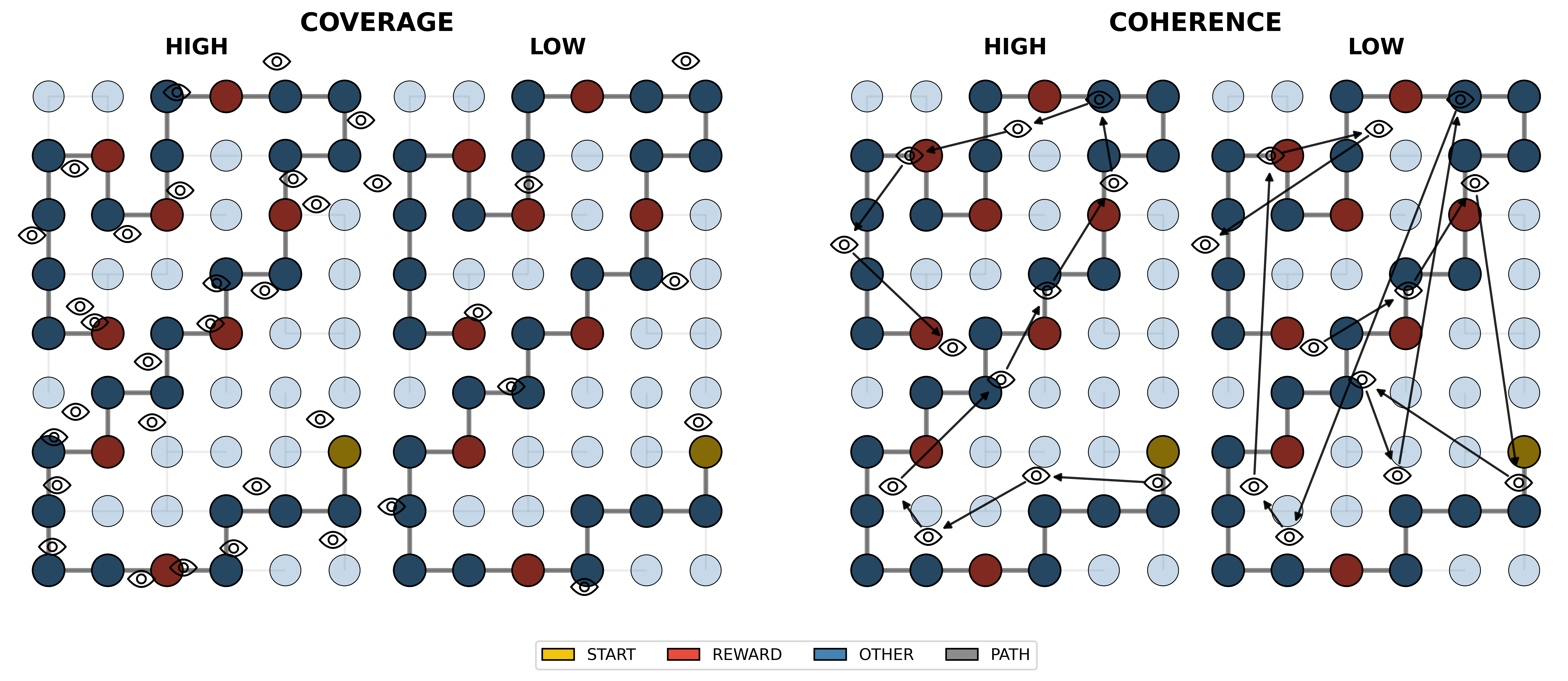}
    \caption{Visual representation of different levels of spatial \emph{coverage} (left) and temporal \emph{coherence} (right). From left to right: high coverage (many of the path nodes are fixated); low coverage (few nodes are fixated); high coherence (transitions between fixations "follow" the path direction, preserving reciprocal ranking within the sequence); low coherence (fixations are on the same nodes as the high coherence case, but their order is different from the one implemented in the path).}
    \label{fig:coverage_coherenceHL}
\end{figure}

For the purpose of comparing visual exploration and path implementation, we projected cursor trajectories into the same space as the fixations by collapsing them into a discrete sequence of selected (or unselected) nodes. In both cases, the sequence of nodes was identified by the identity of the closest node being fixated (or crossed) and by the node’s relative position within the sequence.
Let $\mathcal{P} = (p_1, p_2, \dots, p_N)$ denote a reference path and let $\mathcal{S} = \{s_1, s_2, \dots, s_K\}$ be a sequence of fixations.
We characterized the relation between $\mathcal{S}$ and $\mathcal{P}$ with two main properties: (1) how many of the fixations belonged to the path (spatial coverage), or vice versa, and (2) how many of the transitions between consecutive fixations were present in the path (temporal coherence).

Formally, the spatial coverage (or more briefly, coverage) of $\mathcal{S}$ with respect to $\mathcal{P}$ is:

\begin{equation}
  \mathrm{Coverage}(\mathcal{S}, \mathcal{P})
  = \frac{\mid \mathcal{S} \ \cap \  \mathcal{P} \mid}{\mid \ \mathcal{P} \ \mid},
\end{equation}

i.e.\ the fraction of path nodes being fixated. Note that this definition is not symmetrical, and the coverage of the path provided by the fixations is (in principle) different from the coverage of the fixations provided by the path.

To define the temporal coherence of fixations with respect to the reference path, let us denote the set of ordered couples of nodes $\mathcal{C}(\mathcal{P}) = \{(p_i, p_j)\} _{j> i}$ along the reference path, and $\mathcal{T}(\mathcal{S}) = \{(s_i, s_{i+1})\} _{i = 1}^{k-1}$ the set of transitions between consecutive fixations in a sequence. Note that the reason why we did not compare the transitions between \textit{consecutive} fixations with the transitions between \textit{consecutive} nodes in the path is that gaze does not usually fixate on every single node of the (to be) implemented path. Thus, considering only adjacent nodes as contributing to the order would lead to artificially low values of coherence.

While the direction of the implementation of the path is enforced by a given starting point, the direction of the fixation sequences is (in principle) arbitrary. Thus, when comparing the order of the two sets, we considered both the original sequence of fixations and its reverse. The final temporal coherence (or more shortly, coherence) will be the maximum value between the two comparisons, obtained as:

\begin{equation}
  \mathrm{Coherence}(\mathcal{S}, \mathcal{P})
  =  \frac{\mid \mathcal{T}(\mathcal{S}) \ \cap \ \mathcal{C}(\mathcal{P}) \mid}{\mid \mathcal{T}(\mathcal{S})\mid}
\end{equation}

Similarly to spatial coverage, temporal coherence is also a non-symmetrical property.
\autoref{fig:coverage_coherenceHL} provides a visual representation of high/low coverage/coherence examples.

\subsection*{Clustering fixations into sequences by task phase and spatio-temporal distance} 
\label{sec:clustering}

Although fixations were recorded as a continuous stream, we assumed that, before and during task solution, participants could evaluate different aspects of the task and multiple alternative plans, thus yielding an underlying latent structure that links some fixations into sequences. To identify this putative latent structure, we proceeded in two steps: we first assigned them to four task phases according to the cursor behaviour, as described in \autoref{fig:progression_of_a_trial}. The first phase was the pre-movement period, where participants scanned the elements of the problem. The second was the (forward) execution phase, where participants proceeded with a path implementation. The third consisted of the possible pauses during the execution phase, during which the gaze scanned the map but no path was implemented. The fourth corresponded to the (occasional) backtracks, where participants traced back the path they already implemented, partially, or completely.

This phase-based segmentation provided a coarse temporal description of the behavioural context where fixations occurred within the trial; the clustering procedure described below was then applied within each of these phases to identify finer-grained sequences of fixations.

The need for the refinement of the categorization rests on the assumption that, especially at the beginning and during pauses, multiple plans can be evaluated and gazed at (e.g. partial plans that are discarded; plans to a specific sub-goal or related to distinct parts of the solution; inspection of different map areas, etc.). Because gaze alone does not provide explicit cues about shifts from one putative plan to another, we needed a principled method to infer latent sub‑structure within the fixation stream.

Our goal was to identify groups of fixations that plausibly corresponded to the same local fragment of a plan, while avoiding assumptions about their temporal ordering that could bias the analysis of how sequences of fixations align to executed plans. Fixations were therefore more likely to be clustered together when they occurred close in time and when their corresponding nodes were structurally related within the graph. Structural relatedness was defined by the ease with which two nodes could be connected through the graph: fixations on nodes connected by short paths were considered more likely to belong to the same candidate route fragment than fixations on nodes located in distant or weakly connected regions of the graph. Thus, clustering was based not only on spatial proximity in the visual display but also on the nodes' relationships within the problem structure that participants had to reason about.

To accommodate these properties, we used the generative-embedded Chinese Restaurant Process (geCRP), defined by \cite{Maisto2016}. 
Operationally, the clustering procedure comprised three steps. First, we defined a similarity between fixations that combined their temporal proximity in the scanpath with the structural relation between the corresponding graph nodes. Second, we used the geCRP to estimate which fixations were likely to belong to the same local episode of visual exploration, yielding a co-occurrence structure over fixations. Third, this co-occurrence structure was converted into discrete clusters, grouping fixations that were consistently inferred to belong together. We treated these clusters as fixation sequences -- that is, coherent fragments of visual exploration potentially reflecting the inspection of a route segment, a subgoal, or a local region of the map.

This approach also allowed for the possibility that distinct fragments of visual exploration reflected alternative candidate plans evaluated in parallel (e.g., comparing “turn left” vs. “turn right”); that these fragments partially overlapped (e.g., junction points may mark the end of one partial plan and the beginning of another); and that their internal ordering did not necessarily match the direction of final execution (e.g., a sequence of nodes could be fixated in the backward direction and subsequently traversed in the forward direction).
Technical details of the clustering procedure are provided in the \nameref{sec:supplementary_materials}. The validation procedure and results obtained across synthetic sequence structures are also reported there (\autoref{fig:clustering_validation1}, \autoref{fig:clustering_validation2}, \autoref{tab:kswep}, \autoref{tab:closegroup}, \autoref{tab:bifurcation}, \autoref{tab:longsaw}).

\subsection*{Combining sequences of fixations to reconstruct participants' plans}
Although the clustering method identifies groups of fixations exhibiting some degree of spatio-temporal coherence, no single cluster is expected to capture an entire plan. Planning in this task is often fragmented: participants may inspect different sub‑paths at different times, evaluate alternatives, revisit previously considered options, or interleave plan fragments with unrelated fixations (e.g., to visually check on the remaining time or score). As a result, the structure of the plan is typically distributed across multiple clusters, each capturing only a partial (and potentially noisy) slice of the underlying planning process. To reconstruct a more complete representation of the intended plan, we therefore combined clusters into a larger set of fixation sequences that best matched the subsequently executed path.

We selected this combination $\mathcal{S}^{*}$ by greedily adding sequences that maximized fixation coverage of a reference path, stopping at the smallest set size ($K^{*}$) for which adding a further sequence did not increase coverage (or when all sequences had been exhausted).

Formally, let $\bar{\mathcal{S}} = \{\mathcal{S}_0, ...,\mathcal{S}_M\}$ denote the full set of sequences. A subset of $K$ sequence indices, $\mathcal{I} = \{i,j,...,l\}$ will identify a combination of fixation sequences, denoted by $\mathcal{S}(\mathcal{I}) = \cup _{i \in \mathcal{I}}\mathcal{S}_{i}$. The optimal combination was chosen as follows:

\begin{equation}
  \mathcal{S}^{*}
  = \operatorname*{arg\,max}_{\mathcal{S}(\mathcal{I})}\{\mathrm{Coverage}(\mathcal{\mathcal{S}(\mathcal{I}), \mathcal{P}})\}\quad \text{subject to} \quad \min |\mathcal{I}|
\end{equation}

In cases where combinations containing the same number of sequences provided equal (maximum) coverage, we selected the combination with the highest coherence. Although the optimal coverage $(\mathrm{Coverage}(\mathcal{S}^*, \mathcal{P}))$ is invariant to the order in which sequences are added and to the internal ordering of fixations, these ordering properties become crucial for the coherence of the optimal combination ($\mathrm{Coherence}(\mathcal{S}^*, \mathcal{P})$). 

To extend the coherence definition to the case of combinations of sequences, we accounted for the fact that---as in the single sequence case---the highest coherence could be obtained from either the original sequence $\mathcal{S}_i$, or its reversed version $rev(\mathcal{S}_i)$. Thus, we needed to consider all the possible $2^{K^*}$ combinations of directions for the sequences composing the (best) combination $\mathcal{S}^*$.

Note that since backtrack phases were excluded from the analysis, nodes unselected during backtrack were also removed from the reference path in all comparisons. For the pre-movement phase, the comparison is straightforward and uses the (remaining) selected nodes from the whole implemented solution as reference path.

For pauses and execution phases, the reference path consisted of the upcoming (selected) nodes. However, because these phases could occur after part of the plan had already been implemented, we prevented data leakage by calculating coverage only over the portion of the path that remained to be navigated at the onset of the phase. Thus, during a pause, if a participant fixated primarily nodes that had already been visited, these fixations contributed minimally to the coverage of the upcoming path. Crucially, whereas fixations during pauses and pre-movement phases were compared with a path that had yet to be implemented, fixations during the execution phase were compared with the simultaneous (though asynchronous) sequence of node selection.

\subsection*{Mixed-effects models and mediation analysis}
To quantify the effects of task difficulty and  delay condition on behaviour and gaze, we employed linear mixed‑effects models (LMMs) and generalized linear mixed‑effects models (GLMMs) \citep{Bates2015lme4}. Mixed‑effects models are well‑suited for our design because they account for repeated measurements within participants, allow for participant‑specific variability in baseline levels, and support flexible link functions appropriate for different outcome types (e.g., logit for success rates, log‑count for the number of pauses and backtracks). For each behavioural and gaze-related measure, we modelled fixed effects of difficulty,  delay condition, and their interaction, and included random intercepts for participants to capture between‑participant variability in overall outcome levels. When relevant (e.g., in models of behavioural errors), additional covariates such as the number of available solutions were included to control for residual structural variability across problems.

To evaluate whether certain relationships were direct or instead mediated by changes in gaze behaviour, we complemented these analyses with sequential mixed‑effects mediation models \citep{Imai2010mediation}. In these models, difficulty served as the independent variable, gaze-related quantities (e.g., total number of fixations) as potential mediators, and path‑representation metrics (coverage and coherence) as dependent variables. This framework allowed us to separate the direct influence of difficulty on planning quality from its indirect influence exerted through gaze allocation, while still accounting for participant-level random effects and relevant covariates (such as the length of the reference path). Indirect effects were assessed using the product‑of‑coefficients approach and evaluated with Sobel tests and bootstrap confidence intervals computed via participant‑level resampling. To test differences between distributions we used a one-tailed t-test when normality assumptions held, whereas we preferred the non-parametric Mann--Whitney U test in other cases (e.g., with bimodal distributions). 
 
As an additional control, we assessed the sensitivity of the mediation results to potential unmeasured trial-level factors affecting both the mediator and the outcome. We repeated each mediation analysis while fixing the residual correlation between the mediator and outcome models, $\rho_{\varepsilon}$, over a predefined range and re-estimating the remaining model parameters. For each value of $\rho_{\varepsilon}$, we evaluated the mediator--outcome path. We defined $\rho_{\varepsilon}^{*}$ as the residual correlation at which this path, and consequently the estimated indirect effect, crossed zero. The absolute magnitude of $\rho_{\varepsilon}^{*}$ therefore indicates how strong residual mediator--outcome confounding would need to be to eliminate the estimated indirect effect: values closer to zero indicate greater sensitivity, whereas larger absolute values indicate greater robustness. Sensitivity analyses were conducted for the significant indirect effects involving backtracks, path coverage, and coherence. Further computational details are reported in the \nameref{sec:supplementary_materials}.
Together, these analyses allowed us to examine how planning difficulty was associated with performance both through gaze allocation and independently of it.

\subsection*{Individual differences and clustering tendency}
\label{sec:individual_differences}

To characterize stable individual differences in planning behaviour, we considered delay-adjusted view time (i.e., view time adjusted for the $19\ s $ off-set in the delay condition), spatial coverage, temporal coherence, backtrack density, and pause density. 

For each measure, we first quantified the proportion of variance attributable to between-participant differences using the intraclass correlation coefficient (ICC) \citep{Shrout1979Intraclass} derived from the corresponding mixed-effects model: \begin{equation} ICC = \frac{\tau^2}{\tau^2 + \sigma^2}, \end{equation} where $\tau^2$ denotes the variance of the participant-level random intercepts and $\sigma^2$ the residual variance. 
The ICC quantifies the extent to which variability in each behavioural marker reflects stable differences between participants rather than within-participant variation across trials.

For the subsequent participant-level analysis, we used condition-adjusted values obtained from the mixed-effects models and standardized each measure across participants. Using the Hopkins statistic \citep{lawson1990}, we assessed whether participants tended to form distinct groups within the behavioural space defined by the stable traits identified based on their ICCs. Values close to 0.5 indicate a spatial distribution compatible with the absence of cluster structure, whereas values approaching 1 indicate greater clustering tendency. Finally, we applied principal component analysis (PCA) \citep{Jolliffe2016PCA} to the same standardized measures. The PCA was used to describe the principal dimensions of covariation among participants and to provide a lower-dimensional visualization of individual differences; it was not used to test clustering tendency. We additionally examined the Pearson correlations between participant' component scores and overall success rate.

\section*{Results}

\subsection*{Misleadingness decreases success rate and increases planning demand, whereas imposing a delay period reduces online revision}

\begin{figure}[ht!]
    \centering
    \includegraphics[width=.95\linewidth]{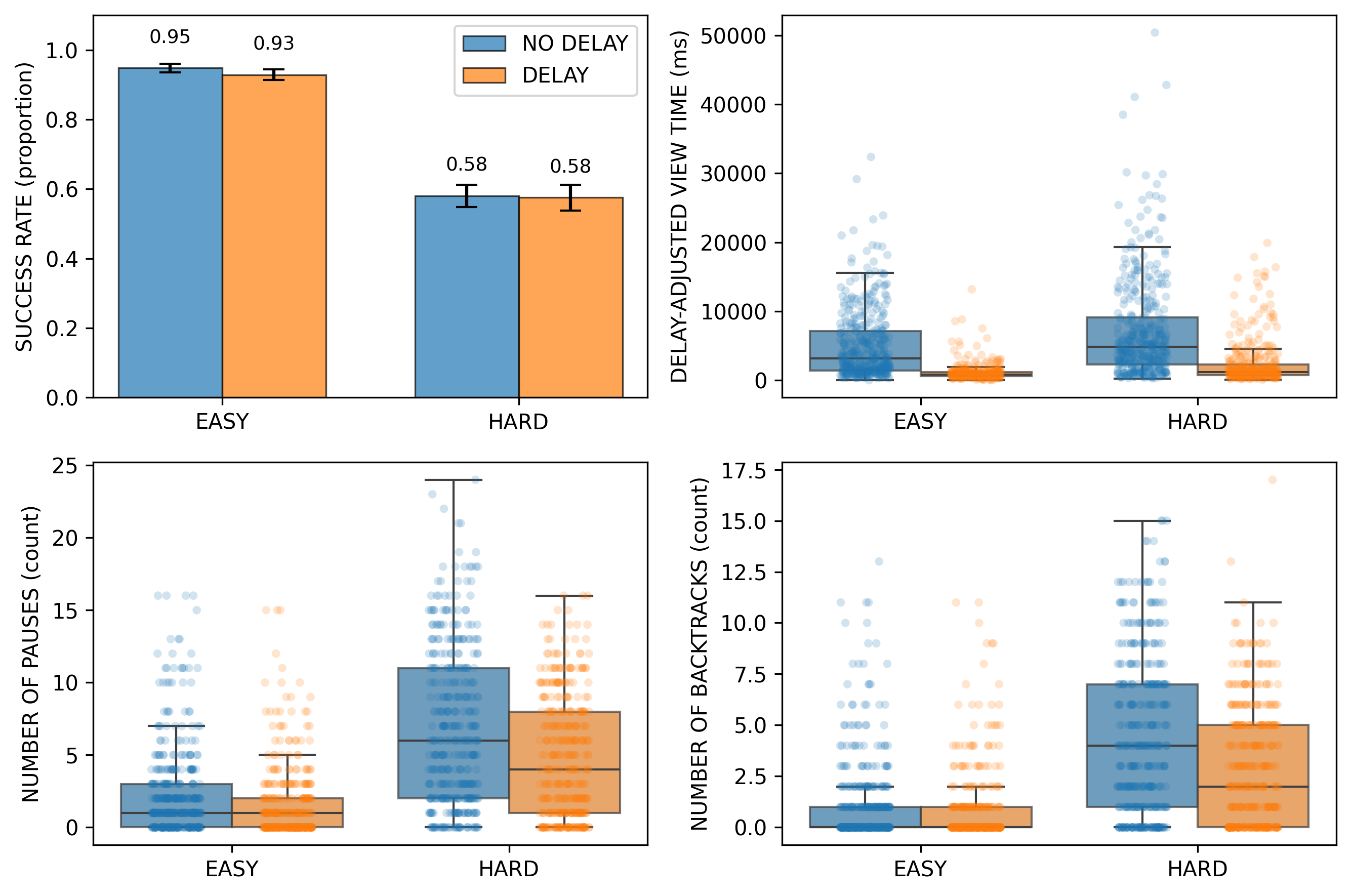}
    \caption{Difficulty comparison across easy and hard problems, split by  delay vs. no-delay conditions. (Top-left) Mean success rate; error bars indicate $\pm 1$ standard error of the mean (SEM), computed across participant means. (Top-right) Delay-adjusted view time box plot. (Bottom-left) number of pauses box plot. (Bottom-right) number of backtracks  box plot. Box plots show the median (horizontal line), interquartile range (box, 25th–75th percentile), and whiskers extending to data points within 1.5 $\times$ inter-quartile range (IQR). Individual participant data points are overlaid with a random horizontal jitter.}
    \label{fig:difficulty_comparison}
\end{figure}

We first  validated misleadingness as a meaningful form of problem difficulty, by testing whether it significantly disrupted participants' performance. As shown in \autoref{fig:difficulty_comparison} (top-left), success rates were lower for hard (high misleadingness) than for easy (low misleadingness) problems. A GLMM confirmed this pattern: difficulty significantly reduced the probability of solving a trial, whereas the  delay manipulation did not reliably affect success (\autoref{tab:success}). We also included the number of solutions as a predictor of success, as this was a property not directly controlled in the problems generation, and found that it had a small positive effect on success; so this residual structural variability between problems contributed only modestly to performance.

Conversely, imposing a delay period did not significantly affect participants' success rate. Our results also show a positive effect of the block -- i.e., whether a trial occurred in the first or the second half of the experiment -- which suggests the presence of some learning process through which participants improved their problem-solving ability over the course of the experiment.  

Furthermore, we tested whether problem difficulty and the delay period affected participants' \textit{delay-adjusted view time}, defined as total view time minus the imposed delay period (19 seconds), if present (\autoref{fig:difficulty_comparison}, top-right). We found that both difficulty and delay affected delay-adjusted view time, although in opposite directions (\autoref{tab:view_time}). Harder problems were associated with longer delay-adjusted view time, suggesting that participants were sensitive to the increased planning demands. Conversely, enforcing a delay period significantly reduced delay-adjusted view time. Finally, we found a significant negative interaction between difficulty and delay, indicating that the difficulty-related increase in view time was attenuated when a delay was imposed.

We also asked whether these changes in delay-adjusted view time affected participants' performance. For this, we conducted a mediation analysis to test whether, and to what extent, delay-adjusted view time mediated the effects of the experimental manipulations on the probability of success (\autoref{tab:mediation_success}). The additional view time associated with harder problems accounted for only a very small proportion of the reduction in success, and its association with performance was weakly negative. Thus, difficulty impaired performance largely independently of how long participants looked before acting. Therefore, although view time reflected participants’ adaptation to task demands, it was not the main factor determining whether they ultimately solved the problem.

\begin{table}[ht]
\centering
\begin{tabular*}{\textwidth}{@{\extracolsep{\fill}}lrrrrl}
\hline
Variables Predicting\\ \textbf{Success Rate} & log-odds & Std.Err & Wald z & p-value & sig\\
\hline\\
Intercept & $1.97$ & $0.33$ & $5.908$ & $3.5\times10^{-9}$ & *** \\
Block & $0.57$ & $0.13$ & $4.315$ & $1.6\times10^{-5}$ & *** \\
Difficulty & $-2.03$ & $0.23$ & $-8.752$ & $<10^{-9}$ & *** \\
 delay & $-0.10$ & $0.13$ & $-0.7322$ & $0.46$ &  \\
Number of solutions & $2.24 \times 10^{-2}$ & $7.9 \times 10^{-3}$ & $2.851$ & $4.4\times10^{-3}$ & ** \\
\hline
\end{tabular*}
\caption{Success Rate---Logistic GLMM (binomial link function). The inclusion of the  delay $\times$ Difficulty interaction was not statistically significant, so the final model discarded this contribution. The p-value significance code: $*** < 10^{-3}$, $10^{-3} < ** < 10^{-2}$, $10^{-2} < * < 0.05$, $0.05 < \text{.} < 0.1$, $> 0.1$.}
\label{tab:success}
\end{table}

\begin{table}[ht]
\centering
\begin{tabular*}{\textwidth}{@{\extracolsep{\fill}}lrrrrl}
\hline
Variables Predicting\\ \textbf{Delay-Adjusted View Time} [ms] & Estimate & Std.Err & t & p-value & sig \\
\hline\\
Intercept & $959$ & $520$ & $1.846$ & $7.31 \times 10^{-2}$ & . \\
Block & $152$ & $210$ & $0.7189$ & $0.4723$ &  \\
 delay & $-3826$ & $300$ & $-12.803$ & $<10^{-9}$ & *** \\
Difficulty & $2052$ & $290$ & $7.033$ & $<10^{-9}$ & *** \\
Difficulty $\times$  delay & $-890$ & $420$ & $-2.102$ & $3.57  \times 10^{-2}$ & * \\
\hline
\end{tabular*}
\caption{Delay-Adjusted View Time ---Linear Mixed Model. Random intercept per participant. The p-value significance code: $*** < 10^{-3}$, $10^{-3} < ** < 10^{-2}$, $10^{-2} < * < 0.05$, $0.05 < \text{.} < 0.1$, $> 0.1$.}
\label{tab:view_time}
\end{table}

\begin{table}[ht]
\centering
\begin{tabular*}{\textwidth}{@{\extracolsep{\fill}}lrrrrl}
\hline
Variables Predicting\\ \textbf{Number of Backtracks} & log-count & Std.Err & Wald z & p-value & sig \\
\hline\\
Intercept & $1.8\times 10^{-1}$ & $9.7 \times 10^{-2}$ & $1.776$ & $7.6 \times 10^{-2}$ & . \\
Block & $-9.3  \times 10^{-2}$ & $6.7 \times 10^{-2}$ & $-1.387$ & $0.16$ &  \\
Difficulty & $1.313$ & $6.9 \times 10^{-2}$ & $19.019$ & $<10^{-9}$ & *** \\
 delay & $-4.23 \times 10^{-1}$ & $7.4 \times 10^{-2}$ & $-5.751$ & $8.9\times10^{-9}$ & *** \\
Delay-adjusted view time & $-1.07 \times 10^{-1}$ & $4.3 \times 10^{-2}$ & $-2.507$ & $0.0122$ & * \\
\hline
\end{tabular*}
\caption{Number of Backtracks -- Negative Binomial GLMM (no interactions). Poisson overdispersion ratio = 2.87. Delay-adjusted view time is a continuous variable, so we standardized it by subtracting the mean and dividing by the standard deviation (SD) of the distribution; its coefficients thus refer to the increase of 1 SD instead of 1 ms. The p-value significance code: $*** < 10^{-3}$, $10^{-3} < ** < 10^{-2}$, $10^{-2} < * < 0.05$, $0.05 < \text{.} < 0.1$, $> 0.1$.}
\label{tab:num_backtrack}
\end{table}

We next examined post-movement onset behaviour, focusing on pauses and backtracks as markers of incomplete or erroneous planning.
In line with the semantics of difficulty, a GLMM estimated that higher difficulty increased the number of backtracks, meaning that more misleading problems required greater online revision and corrections (\autoref{tab:num_backtrack}). In contrast, longer pre-movement view time was associated with fewer backtracks. Thus, although additional view time did not increase the probability of solving the problem, it was associated with smoother execution once movement had begun.

Because we observed that harder problems were associated with longer view times we explored the mediation path linking task difficulty, delay-adjusted view time and number of backtrack.
Harder problems were associated with a more intense online revision, but also with longer view times which in turn mitigated this effect (\autoref{tab:mediation_num_backtrack}). A sensitivity analysis (see \nameref{sec:supplementary_materials}) showed that the indirect effect crossed zero at $\rho_{\varepsilon}=-0.102$, indicating that a residual mediator–outcome correlation of this magnitude would be needed to explain away this mediation effect.
Similarly, enforcing a time delay also (directly) decreased the number of backtracks, although delay-adjusted view time was itself reduced in this condition. A potential (and somewhat trivial) confound is that imposing a  delay period leaves less time in which backtracks can occur. However, a control using an LMM on the density of the number of backtracks (over the remaining available time) confirmed that the effect of the delay condition retained its significance (\autoref{tab:num_backtrack_per_time}).

\begin{table}[ht]
\centering
\begin{tabular*}{\textwidth}{@{\extracolsep{\fill}}lrrrrl}
\hline
Variables Predicting\\ \textbf{Number of Pauses} & log-count & Std.Err & Wald z & p-value & \\
\hline\\
Intercept & $9.29 \times 10^{-1}$ & $9.2 \times 10^{-2}$ & $10.104$ & $<10^{-9}$ & *** \\
Block & $-2.65\times 10^{-1}$ & $5.3 \times 10^{-2}$ & $-4.980$ & $6.3\times10^{-7}$ & *** \\
Difficulty & $1.099$ & $5.4 \times 10^{-2}$ & $20.202$ & $<10^{-9}$ & *** \\
 delay & $-4.26 \times 10^{-1}$ & $5.9\times 10^{-2}$ & $-7.289$ & $<10^{-9}$ & *** \\
Delay-adjusted view time & $-4.6\times 10^{-2}$ & $3.3\times 10^{-2}$ & $-1.398$ & $0.1620$ &  \\
\hline
\end{tabular*}
\caption{Number of Pauses -- Negative Binomial GLMM (no interactions). Poisson overdispersion ratio = 3.39. The p-value significance code: $*** < 10^{-3}$, $10^{-3} < ** < 10^{-2}$, $10^{-2} < * < 0.05$, $0.05 < \text{.} < 0.1$, $> 0.1$.}
\label{tab:num_pauses}
\end{table}

Finally, this pattern of results is partially shared with the GLMM description of the number of pauses (\autoref{fig:difficulty_comparison} bottom-left; \autoref{tab:num_pauses}). The first significant difference is that progressing from the first to the second half of the experiment reduces the number of pauses, suggesting a learning effect---already shown in the success rate analysis. By contrast, the effect of delay-adjusted view time is not significant.
As in the case of backtracks, we ruled out the possible confound of the (smaller) available time in the  delay condition by fitting an LMM on the density (over time) of the number of pauses, which showed that the effect of the delay condition remained (highly) significant (\autoref{tab:num_pauses_per_time}).

Overall, these results support the interpretation of misleadingness as a measure of problem difficulty: more misleading problems were less likely to be solved and required more online revisions (pauses and backtracks). On the other hand, imposing a delay period (as well as increasing view time) does not change success rate, but redistributes the planning effort, reducing the need for online revision.

\FloatBarrier
\subsection*{Task phase, difficulty and delay shape the distribution of fixations to start, reward and other nodes}

\begin{figure}
    \centering
    \includegraphics[width=.9\linewidth]{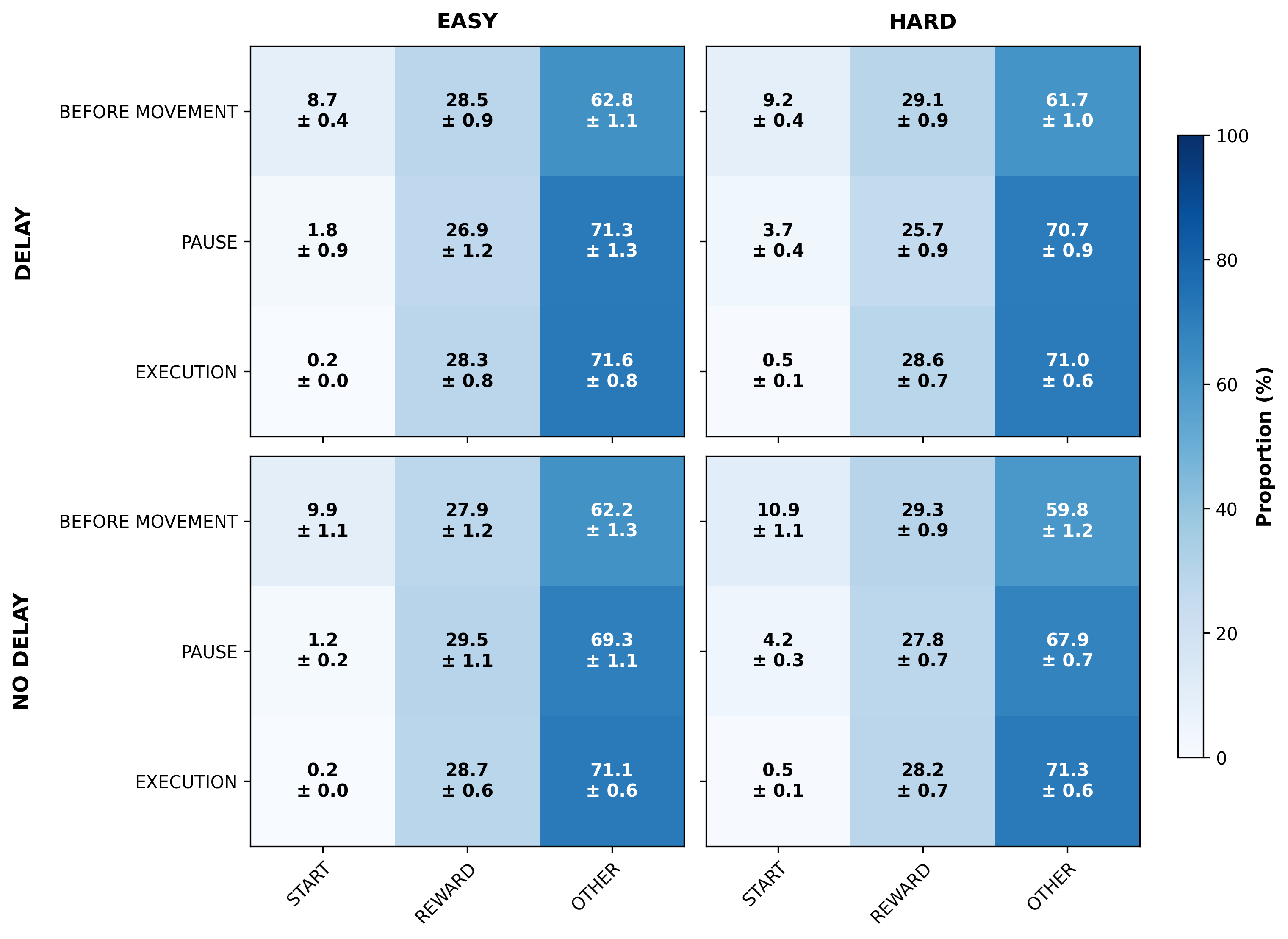}
      \caption{Fixation distribution across navigation phases, node types, and task conditions. Each panel shows the percentage of fixations directed to different node types (columns: Start, Reward, Other) during three distinct navigation phases (rows: Before Movement = pre-movement planning, Pause = mid-execution pauses, Execution = forward navigation). The four panels represent all combinations of  delay condition (rows:  delay vs No- delay) and difficulty level (columns: Easy vs. hard). Color intensity represents the proportion of fixations ($0-100\%$), with darker blue indicating higher fixation proportions. Values are normalized within each phase-difficulty- delay combination, summing to 100$\%$ across the three node types. Errors are the standard error of the mean across all trials.}
    \label{fig:fixations_proportions_per_phase}
\end{figure}

To gain further insight on participants' visual exploration of the problems, we next quantified the distribution of fixations directed to each node category -- start, reward, or other (non-reward) nodes -- across the three task phases -- before movement, during pauses, and during execution -- and across difficulty and delay conditions, see \autoref{fig:fixations_proportions_per_phase}. Fixation proportions were modelled using a GLMM with phase, difficulty,  delay, and all their interactions as fixed effects, and a random intercept for participant (\autoref{tab:glmm_start}, \autoref{tab:glmm_reward}, \autoref{tab:glmm_other}).
We then conducted post‑hoc analyses based on estimated marginal means to test ordered contrasts across phases (\autoref{tab:emmeans_contrasts}).

Across every combination of difficulty and delay conditions, the percentage of fixations directed to the starting node satisfied the chain of inequalities Before movement > Pause > Execution.
Fixations on the other (non‑reward) nodes showed the inverse ordering: Before movement < Pause < Execution (except in the {delay} condition, where fixation proportions during pauses and execution did not differ significantly).

In contrast, fixations on reward nodes showed a far more stable pattern across phases and conditions. For hard problems, however, they were lower during pauses than before movement both with and without delay, and lower during pauses than during execution when a delay was imposed.

These contrast analyses highlight a substantial shift in gaze allocation while proceeding from the planning phase before movement to the execution phase: the amount of fixations on the starting node progressively shifts towards the non-reward ("other") nodes. It is worth noting that fixations on reward nodes during the pre‑movement phase are disproportionately frequent relative to their structural prevalence in the map solution. For the shortest solutions (ranging from 32 to 35 nodes across problems), the expected reward‑to‑other ratio is roughly 1:3 -- a lower bound, since all longer solutions increase this imbalance -- yet in the pre‑movement phase the observed ratio is closer to 1:2. During execution, fixation proportions approach those expected from uniform sampling along the solution path. Pauses tend to fall between these two extremes (except for the fact that when a delay is imposed, rewards are fixated significantly less than in the other phases).

The analysis of the distribution of fixations across node types must be complemented by a quantification of the extent to which individual nodes are repeatedly fixated. Indeed, it is possible for two fixation sequences to have exactly the same distribution across node types, while differing substantially in the diversity of the individual nodes they fixate. For example, a visual exploration strategy that first directs fixations towards the subgoal to be achieved and then identifies intermediate nodes to construct a complete solution could yield the same type-level statistics as a visual exploration strategy that instead repeatedly samples the same partial plan.

\begin{figure}[ht!]
    \centering
    \includegraphics[width=1\linewidth]{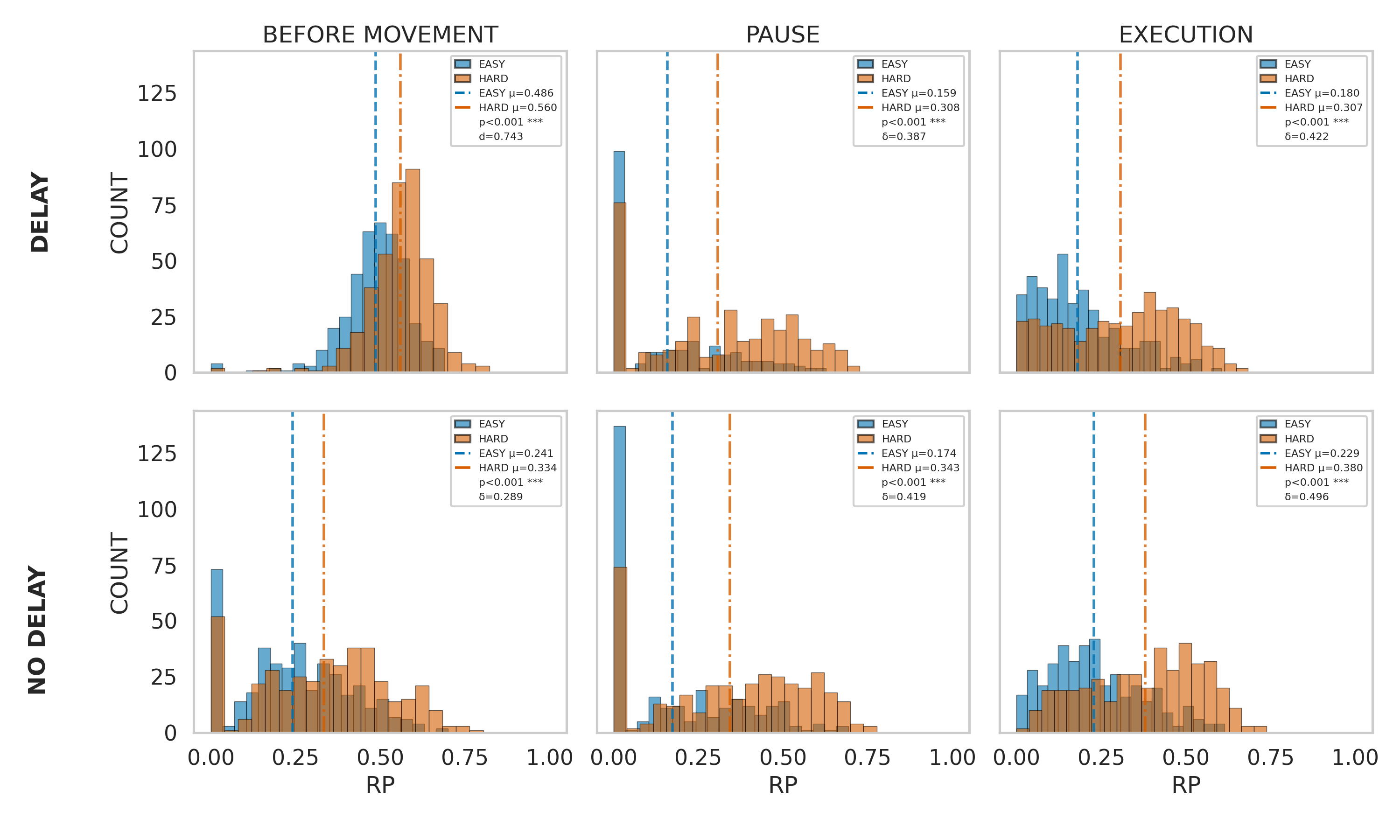}
    \caption{Histograms showing the distribution of re-fixation proportions split by  delay condition (rows), phase (columns), and task difficulty (colors). Blue represents easy trials, orange represents hard trials. Dashed vertical lines indicate mean values (---for Easy, -. for Hard). Each legend reports mean values and statistical test results comparing easy vs. hard difficulty: t-test for Before Movement/ delay condition, Mann--Whitney U test for all other conditions. Significance codes: *** p<0.001, ** p<0.01, * p<0.05, . p<0.1, ns p>0.1. Effect sizes are reported using a Cohen's $d$ for the Before Movement/Delay condition; a Cliff's $\delta$ was used for all the other (strongly non-Gaussian) cases.}
    \label{fig:oversampling_Repeat Ratio}
\end{figure}

We quantified this behaviour by measuring the re-fixation proportion (RP).
The distributions shown in \autoref{fig:oversampling_Repeat Ratio} revealed significant differences between easy and hard problems: across all combinations of delay condition and task phase, the $RP$ was substantially higher for harder problems.
Consistent with the planning–revision trade-off observed in the analyses of backtracking and pauses, enforcing a delay period resulted in a higher proportion of re-fixations before movement; during execution, however it led to less repetitive gaze behaviour.

A potential concern with interpreting these results is that harder problems elicited more total fixations. Since the solutions to all problems contain approximately the same number of relevant nodes, an increase in the number of fixations, while the number of unique fixated nodes remains bounded, would inflate the re-fixation proportion, even if participants’ re-fixation behaviour itself did not truly change. In other words, the higher $RP$ observed for difficult problems could mechanically arise from more fixations distributed over a comparable number of nodes, rather than reflecting a genuine shift in visual exploration.

To address this possibility, we conducted a control analysis in which we fitted an LMM that included the total number of fixations as a covariate to estimate the effects of difficulty and  delay condition on $RP$ separately for each phase (\autoref{tab:logit(Repeat Ratio)}). Crucially, the effect of task difficulty on $RP$ remained significant in all phases and conditions (although the Difficulty $\times$ Delay interaction indicated that the before-movement effect was attenuated under delay), indicating that the increase in re-fixations for harder problems cannot be fully accounted for by a larger number of fixations. 

Together, these results indicate that fixation patterns across node types (start, reward, and other nodes) change significantly across task phases, suggesting a transition from higher-level planning directed primarily towards start and reward nodes before movement to lower-level planning directed towards other nodes connecting them during execution. Greater difficulty increases refixations of the same nodes, plausibly reflecting the greater demands of finding and maintaining a solution. In contrast, the delay period increases the proportion of refixations before movement but decreases it during execution, consistent with the idea that planning effort is shifted towards the pre-movement period.

\subsection*{Reconstructing the upcoming path from visual fixation sequences}

The analyses above showed that before movement, problem difficulty and delay affected both the distribution of fixations across node types and the tendency to revisit previously inspected locations. However, these aggregate measures are insufficient to assess which fixations, or sequences of fixations, are associated with the subsequently executed plan. 

To assess more directly the alignment between gaze before movement and subsequent plan execution, we clustered fixations into coherent sequences (see \nameref{sec:clustering}) and identified the smallest combination of sequences that maximized spatial coverage of the subsequently executed path. Fixations were excluded from this combination when including their sequence did not improve spatial coverage or temporal coherence, either because the sequence was redundant, because it reduced temporal coherence without improving coverage, or because it increased sequence length without improving either measure. Then, we assessed various properties of the reconstruction of the executed path, including the proportion of fixations selected, their distribution across path-relevant and out-of-path nodes, and the alignment of sequence direction with the upcoming path.

As a preliminary characterization of the clustering output, we examined the number and length of the inferred fixation sequences across task phases and experimental conditions (\autoref{fig:sequence_distribution}).  delay was associated with a greater number and longer sequences before movement, whereas difficulty had little effect in this phase. During pauses, however, harder problems elicited more sequences without reliably changing their length.

We next characterized the composition of the selected fixation sequences, as this can reveal how visual exploration supports planning; for example, how broadly participants explore easy and difficult problems, whether they repeatedly sample the same sequences, and whether visual sampling proceeds in the same or opposite direction to path execution.

\begin{figure}
    \centering
    \includegraphics[width=1\linewidth]{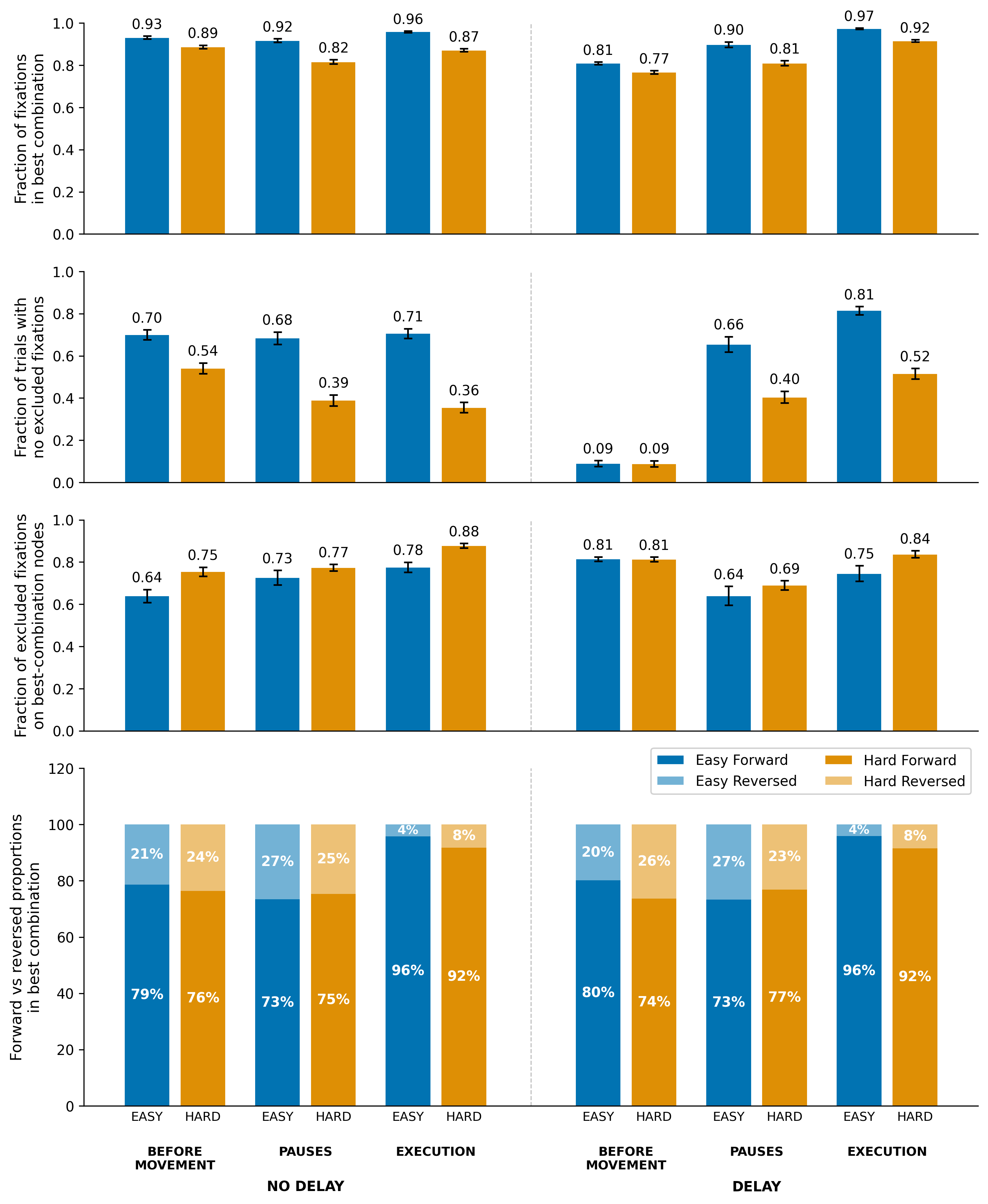}
    \caption{Properties of the upcoming plan reconstruction. First row: Fraction of fixations that were included in the best combination, for each phase. Error bars are the standard error of the mean for the aggregated trials. Second row: Fraction of trials in which no exclusion occurred (i.e., every fixation sequence was selected into the best combination, so no fixations were excluded). Third row: Fraction of the excluded fixations -- those belonging to sequences not selected into the best combination -- that nonetheless landed on nodes that were part of the best combination; computed only over the trials that actually had excluded fixations, with trials showing no exclusion ignored rather than counted as zero. Fourth row: Proportion of fixations aligned with the optimal direction, relative to the reference path direction, across the fixation sequences constituting the best combination.}
    \label{fig:fraction_and_direction}
\end{figure}

The aggregated results are shown in \autoref{fig:fraction_and_direction} and were compared across task phases, difficulty levels, and  delay conditions (contrasts reported in \autoref{tab:combined_difficulty_contrasts}). We found that across task phases and delay conditions, the proportion of fixations included in the selected combination was slightly lower for harder problems, meaning that a larger proportion of fixations was excluded from the best combination (\autoref{fig:fraction_and_direction}, first row). Excluded fixations included repeated sampling of locations on the upcoming path, potentially reflecting verification or consolidation, as well as sampling outside that path. Furthermore, the fraction of trials with no exclusions -- that is, in which all fixation sequences were included in the reconstruction -- was greater for easy than difficult problems (\autoref{fig:fraction_and_direction}, second row), except before movement when a delay was imposed, where only a very small fraction of trials showed no exclusions.

Focusing instead on trials with at least one excluded fixation sequence, we measured the proportion of excluded fixations directed towards nodes represented in the best-fitting combination (\autoref{fig:fraction_and_direction}, third row). 
More than half of these excluded fixations were nevertheless directed towards path-relevant locations, potentially indicating repeated sampling of the selected path for plan verification or consolidation. This proportion was numerically higher for hard problems across most phases and delay conditions, but the difference was significant only before movement and during execution in the no-delay condition.

Finally, across all task phases and delay conditions, the sequences selected to reconstruct the upcoming path predominantly followed the direction in which the path was subsequently executed (\autoref{fig:fraction_and_direction}, fourth row). Nevertheless, before movement and during execution, hard problems showed a slightly larger proportion of selected sequences that traversed path nodes in the reverse order.

Together, these results indicate that, before movement, visual exploration of harder problems is broader, especially in terms of repeated sampling of the upcoming plan -- potentially reflecting verification or consolidation. Although forward sampling predominates, difficult problems are associated with a less consistently forward organization of path-related gaze.

\FloatBarrier
\subsection*{Assessing the alignment between visual fixation sequences and the upcoming path: spatial coverage and temporal coherence} 

We next asked how well the selected subset of fixation sequences identified in the previous analysis captured the structure of the upcoming path, across task phases (before movement, pauses, execution) and conditions (difficulty and delay). We focused on two complementary measures: \emph{spatial coverage}, indexing the proportion of the upcoming path that was sampled, and \emph{temporal coherence}, indexing how closely fixation order matched the order of subsequent actions. 

When assessing how coverage and coherence varied with difficulty and delay conditions, we considered a potential fixation-count confound. Hard problems and trials in the delay condition were associated with more fixations, thereby providing more opportunities to fixate nodes belonging to the upcoming path. Greater visual exploration could therefore mechanically increase spatial coverage and apparent temporal coherence, even without a corresponding change in the organization of path-related gaze. We used mediation analyses to distinguish indirect associations through total fixation count from associations that remained after accounting for fixation count.(\autoref{tab:mediation_waiting}; \autoref{tab:mediation_no_waiting}; \autoref{tab:critical_rho_eps_mediation_excl_p19}). 

\begin{figure}
    \centering
    \includegraphics[width=\linewidth]{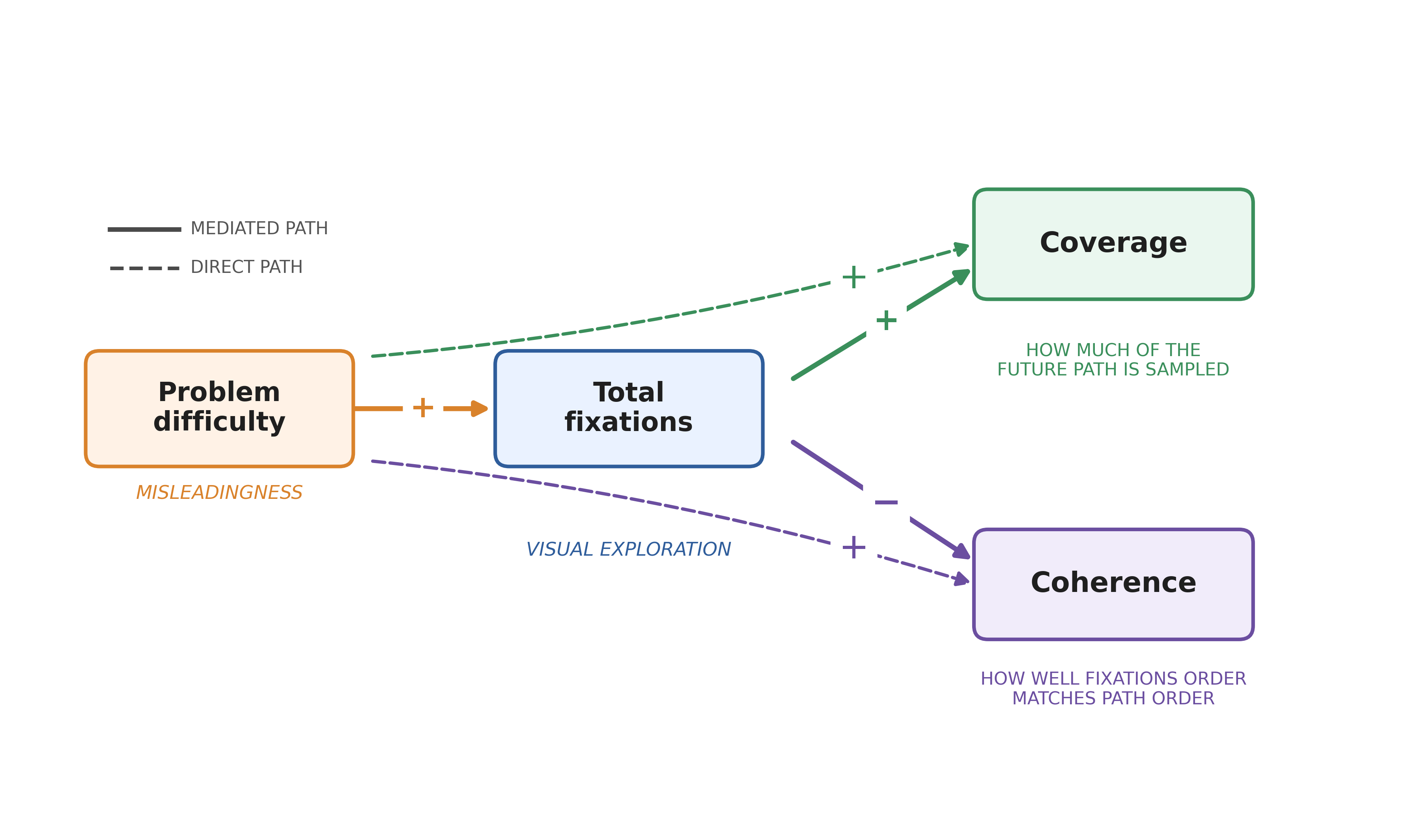}
    \caption{General scheme representing the mediation analysis results. (When significant) Problem difficulty is associated with a higher value of both spatial coverage and temporal coherence, thus a higher-quality plan representation. However, harder problems are also associated with more fixations, which improve path coverage but do so in a less orderly way. This pattern is shared across all phases, though not always significant.}
    \label{fig:mediation_diagram}
\end{figure}

\autoref{fig:mediation_diagram} summarizes the directions of the direct and indirect effects estimated in the mediation analyses. As established above, harder problems elicited more fixations. Significant direct effects of difficulty on path coverage and temporal coherence were positive. By contrast, the indirect effects mediated by fixation count differed between the two measures: higher fixation counts were associated with greater coverage but lower coherence. Thus, greater difficulty was associated with gaze sampling a larger proportion of the upcoming path, partly because it elicited more visual exploration; however, the increase in fixation count was also associated with lower temporal coherence.

Although the direction of these effects was broadly consistent across task phases and delay conditions, their magnitude and statistical significance varied. Before movement and during pauses, the greater coverage observed for harder problems was largely mediated by the increased number of fixations. This indirect effect was significant in both delay conditions.

During execution, by contrast, difficulty predicted greater coverage even after accounting for the number of fixations. This pattern is consistent with gaze being more selectively directed towards upcoming path elements, potentially reflecting more efficient monitoring of upcoming actions. A significant residual  indirect effect was present only in the delay condition.

Sensitivity analyses of the significant indirect effects on coverage showed that $|\rho_{\varepsilon}^*| > 0.5$ in every combination of task phase and delay condition (\autoref{tab:critical_rho_eps_mediation_excl_p19}).

For temporal coherence, the (positive) direct and (negative) indirect effects of difficulty were significant only in the delay condition (\autoref{fig:mediation_diagram}). When no delay was enforced, only the direct effect during execution retained its significance. The robustness of the indirect effects to unmeasured mediator--outcome confounding varied across task phases: the critical residual correlation exceeded $0.2$ before movement onset and during execution, but was lower during pauses ($|\rho_{\varepsilon}^*| = 0.144$; \autoref{tab:critical_rho_eps_mediation_excl_p19}). The indirect effects on temporal coherence should therefore be interpreted more cautiously, particularly during pauses.

\begin{figure}
    \centering
    \includegraphics[width=1\linewidth]{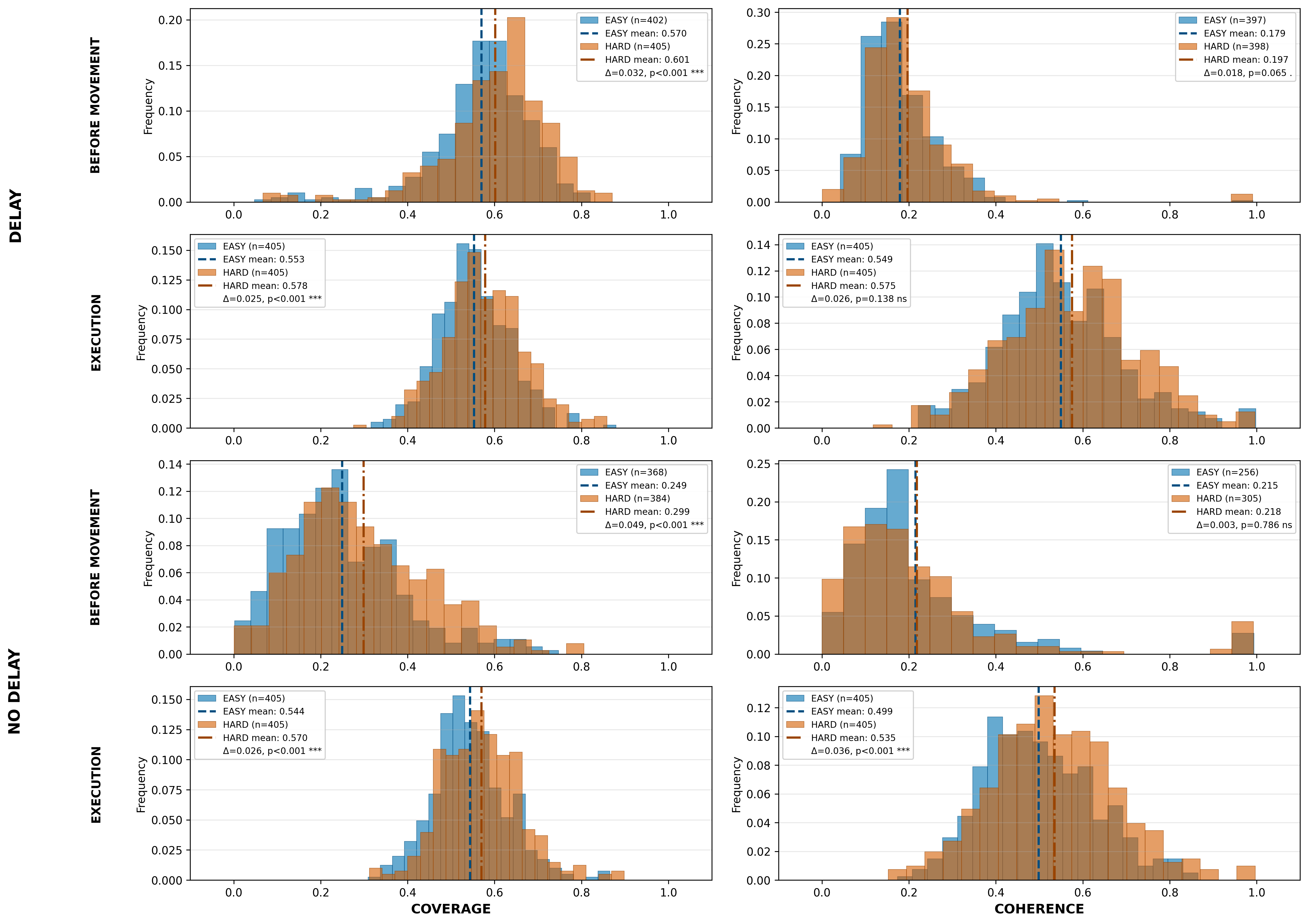}
    \caption{Distribution of spatial coverage (left) and temporal coherence (right) of the best fixation-sequence combination across trial phases, difficulty and delay conditions. The values are adjusted by the random effects (participants' random intercept) and by the effect of the reference path length. The top two rows show the  delay condition; the bottom two rows show the no- delay condition. Within each group, rows correspond to the before-movement and execution phases. Histograms show the distribution for easy (blue) and hard (orange) trials; dashed and dash-dotted lines indicate group means. Each legend reports sample size, mean values, mean difference ($\Delta$), the p-value corresponds to the total effect reported from the mediation analysis. Note that, when estimating temporal coherence, we excluded best combinations of sequences containing fewer than four fixations, as they tended to produce inflated coherence values while providing an insufficiently informative description of the forthcoming path.}
    \label{fig:cov_coh_adjusted}
\end{figure}

Further insight into the effects of difficulty and delay on spatial coverage and temporal coherence can be obtained by examining their fixation-count-adjusted distributions (\autoref{fig:cov_coh_adjusted}). The distributions of both spatial coverage and temporal coherence show greater median values for hard than easy problems across most combinations of task phase and delay condition, although the difference reaches significance only for some combinations. Furthermore, in the no-delay condition, both spatial coverage and temporal coherence show different distributions before movement and after execution, with significantly greater values during execution. When a delay is imposed, the pre-movement distribution of spatial coverage shifts significantly towards that observed during execution, whereas the distribution of temporal coherence does not.

Together, these findings highlight contrasting effects of task difficulty on two aspects of gaze-based planning. Hard problems were associated with visual exploration of a larger proportion of the upcoming path, and this greater coverage was partly mediated by increased fixation count. However, higher fixation counts were also associated with lower temporal coherence. Thus, under higher task demands, gaze covered more of the upcoming path, but the order of fixations corresponded less closely to the order in which the associated actions were subsequently executed.
\FloatBarrier

\subsection*{The analysis of individual differences highlights a planning–revision continuum rather than discrete behavioural types}

Finally, we aimed to assess whether individual differences reflected discrete planning strategies or instead varied along a continuum. For this, we performed an intraclass correlation coefficient (ICC) analysis using delay-adjusted view time, spatial coverage, temporal coherence, backtrack density, and pause density (see \nameref{sec:individual_differences}).

The intraclass correlation coefficient (ICC) analysis indicated that delay-adjusted view time had the highest between-participant reliability ($ICC = 0.30$), followed by spatial coverage ($ICC = 0.19$). Backtrack density ($ICC = 0.12$) and pause density ($ICC = 0.10$) showed lower between-participant reliability. Temporal coherence showed little stable between-participant variation ($ICC = 0.05$) and, given its low ICC, was excluded from the subsequent participant-level analysis. 

\begin{figure}[ht]
    \centering
    \includegraphics[width=0.65\linewidth]{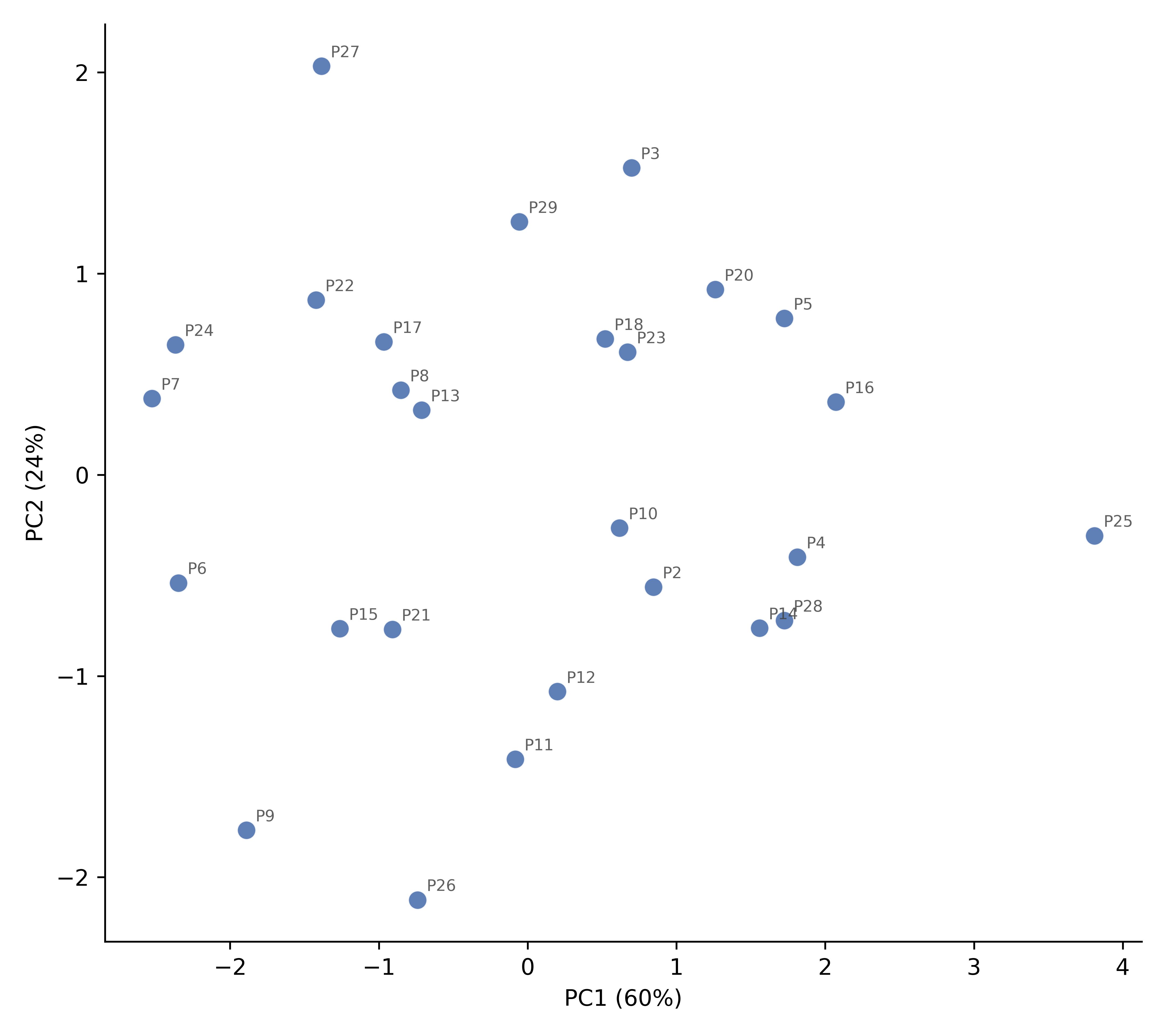}
    \caption{Participants' projections in the space of the first two principal components.}
    \label{fig:pca}
\end{figure}

The Hopkins statistic, computed in the four-dimensional space defined by delay-adjusted view time, path coverage, backtrack density, and pause density, was $0.56 \pm 0.04$. Its proximity to the value expected under spatial randomness ($0.5$) provided little evidence that participants formed distinct behavioural clusters. PCA was therefore used to characterize continuous dimensions of individual variation (\autoref{tab:pc_coef_levels}). The first principal component explained 60\% of the variance and loaded positively on delay-adjusted view time and spatial coverage and negatively on backtrack and pause densities. We interpreted this component as capturing a planning-revision dimension. The second component explained a further 24\% of the variance and loaded primarily on pause density. Together, the first two components accounted for 84\% of the variance. The distribution of participant scores on these components showed no evident separation into discrete groups (\autoref{fig:pca}). Higher scores on the first component were associated with greater overall success, whereas higher scores on the second component were associated with lower success. 

Despite reporting the correlations between the planning-revision components and success, we do not interpret them because these components include pauses and backtracks, whose relationship with task success may be partly definitional. We therefore separately examined whether spatial coverage and temporal coherence were associated with participants' overall success rates. At the participant level ($N = 27$), neither spatial coverage nor temporal coherence was significantly associated with success (coverage: $r = 0.25$, $p = 0.208$; coherence: $r = 0.07$, $p = 0.726$). 

Together, our results suggest that the observed individual differences were more consistent with continuous variation -- particularly in the balance between pre-movement planning and online revision -- than with discrete behavioural types. Furthermore, our results provide no evidence that greater pre-movement coverage or coherence translated directly into higher overall success, although the small sample of participants limits the strength of this conclusion.

\begin{table}[t]
\centering
\begin{tabular*}{\textwidth}{@{\extracolsep{\fill}}lcccc@{}}
\toprule
Feature & PC1 & PC2 & PC3 & PC4 \\
\midrule
View-time level          & $+0.54$ & $+0.43$ & $+0.13$ & $+0.71$ \\
Coverage level           & $+0.56$ & $+0.31$ & $+0.36$ & $-0.68$ \\
Backtrack density level     & $-0.53$ & $+0.26$ & $+0.80$ & $+0.11$ \\
Pause density level         & $-0.34$ & $+0.81$ & $-0.46$ & $-0.15$ \\
\midrule
Explained variance       & 60\% & 24\% & 10\% & 6\% \\
Correlation with success & $+0.42^{*}$ & $-0.63^{***}$ & $-0.06$ & $-0.16$ \\
\bottomrule
\end{tabular*}
\caption{PCA of the four condition-adjusted $z$-scored behavioural markers. All four components are reported. Entries are the unit-norm eigenvector weights defining each component; signs are arbitrary up to a joint flip within a component. The last row gives the Pearson $r$ of each component score with overall success, uncorrected. $^{*}\,p<.05$; $^{**}\,p<.01$; $^{***}\,p<.001$.}
\label{tab:pc_coef_levels}
\end{table}

\FloatBarrier
\section*{Discussion}

Despite extensive research on human planning, its covert nature and flexibility in response to task structure and situational constraints leave several fundamental questions unresolved. Here, we asked how detailed planning is before movement, as reflected in visual exploration of the problem; to what extent visual exploration aligns spatially and temporally with the executed plan and whether visual exploration is modulated by task factors, such as problem difficulty and temporal constraints, as well as by individual differences.

To address these questions, we used a problem-solving task in which participants had to reach multiple targets on a grid without visiting the same node twice, while their actions and gaze were recorded throughout the task. We manipulated two factors: planning difficulty, operationalized as \textit{misleadingness}, or the mismatch between nodes’ spatial proximity in the graph and their relative distance along the solution path; and  delay, by allowing participants either to act immediately or forcing them to wait before initiating movement.

First, we assessed whether our behavioral manipulations influenced participants' performance. We found that more misleading problems were associated with lower success rates and more frequent pauses and backtracks during execution, validating misleadingness as an index of difficulty. Notably, this newly introduced construct predicted difficulty above and beyond structural properties of the problems identified in previous studies, such as solution length, number of subgoals, number of turns, and number of solutions \citep{Giannopoulos2014, Raubal2002,Eluchans2025AdaptivePlanning, Wiener2003, Wiener2009, Kaller2012, Proctor2018, simonelli2025structuring, simonelli2026foraging}. Misleadingness quantifies how strongly the shortest plan diverges from the problem's perceptual layout, particularly the physical proximity of nodes. Thus, high misleadingness values expose a limitation of greedy planning strategies that implicitly assume “close in space, close in the solution”. When spatial proximity no longer predicts proximity within the solution sequence, search slows, and planners may need to expend more effort to maintain the correct ordering -- an interpretation consistent with both longer view times and more frequent re-fixations. The cost of deviating from a perceptual prior aligns with theories of bounded rationality that posit a default policy -- understood as habitual behaviour or a behavioural prior -- and define the cognitive cost of planning as the cumulative deviation from this default \citep{rubin2012trading, piray2020linear, todorov2006linearly, lancia2023humans}. The effectiveness of misleadingness is also consistent with hierarchical planning models that rely on clustering based on perceptual proximity \citep{pezzulo2018hierarchical, ribas2011neural, Solway2014, tomov2020discovery, donnarumma2016problem, vanDijkSubgoal}, as high misleadingness would make such clustering less effective.

Conversely, imposing a delay did not change success rates but reduced pauses and backtracks during execution. Performance also changed over the course of the experiment independently of the experimental manipulations. Participants were more likely to solve problems successfully and made fewer pauses in the second half of the experiment, consistent with adaptation to the task through experience. However, backtracks did not reliably decrease in the second half, suggesting that experience may not affect all aspects of behaviour equally.

Next, we asked whether our experimental manipulations affected the amount of visual exploration before movement and whether this was related to subsequent performance. We found that participants visually explored harder problems for longer, particularly when no delay was imposed. However, neither additional view time nor the imposed delay reliably improved success. Instead, both were associated with fewer backtracks, while the delay also reduced pauses. Additional visual exploration may therefore have helped participants consolidate their plans and execute them more smoothly, without necessarily enabling them to overcome the structural difficulty of misleading problems. This dissociation suggests that successful solution discovery and efficient plan execution capture distinct aspects of planning quality: additional visual exploration may stabilize action selection and reduce the need for online revision, even when it does not increase the likelihood of finding the correct solution.
 
Furthermore, we asked whether and how the content of visual exploration changes across task phases and as a function of our experimental manipulations. We found that the targets of visual exploration changed systematically across task phases. Before movement, gaze was relatively concentrated on the starting node and disproportionately directed towards reward nodes relative to their prevalence on the solution path. During execution, gaze shifted towards intermediate nodes and more closely reflected the composition of the upcoming path. Gaze during pauses generally showed an intermediate distribution, although this pattern varied across conditions. These findings are consistent with a transition from the early prioritization of goal-relevant locations to visual exploration that is more closely coupled to the current action sequence.

Difficulty and delay produced more selective changes within this broader phase-dependent organization. Difficulty altered fixation allocation differently across phases, indicating that its effects on visual exploration depended on whether participants were preparing, reconsidering, or executing a plan. Furthermore, difficulty increased re-fixations across all task phases, even after controlling for the total number of fixations. Harder problems therefore affected not only the amount of visual exploration but also its iterative character, leading participants to return more frequently to previously inspected locations. Such revisiting may reflect the comparison, verification, or consolidation of candidate route segments, consistent with iterative and approximate accounts of planning \citep{binder2025humans,correa2023humans,huys2015interplay}.

The delay manipulation instead redistributed sampling across phases. Most notably, in hard problems, when movement was delayed, reward nodes were sampled less during pauses than before movement or during execution. This may indicate that pauses following an extended preparation period were less strongly oriented towards identifying or reassessing goal locations and more towards processing the intermediate structure needed to continue execution. The delay also increased re-fixations before movement but reduced them during execution. Together with the reductions in pauses and backtracks, this pattern is consistent with the delay increasing visual exploration of the upcoming path before movement, and hence potentially its level of detail, thereby reducing the need for replanning during execution.

Our next analysis examined how visual exploration prior to movement aligns with subsequent path execution. We found that, in harder trials, a smaller proportion of fixations contributed to the reconstruction of the upcoming path. Excluded fixations included mostly repeated sampling of locations on the upcoming path, potentially reflecting verification or consolidation.

We next examined whether the fixation sequences selected to reconstruct the upcoming path tended to follow the same direction as subsequent execution. Across phases and conditions, most selected sequences progressed in the forward direction of the upcoming path, indicating that path-related gaze often preserved the direction of the subsequent action sequence. A minority progressed in the reverse direction, showing that visual exploration of upcoming path elements did not exclusively unfold from earlier to later actions. Such reverse sampling may reflect the evaluation of connections between route segments or candidate subgoals. The balance between forward and reverse sampling varied across task phases and difficulty levels. Harder problems were associated with a lower proportion of forward-directed sequences before movement when a delay was imposed and during execution with and without a delay, whereas no reliable difficulty differences emerged during pauses. Thus, although forward sampling predominated, difficult problems sometimes produced a less consistently forward organization of path-related gaze. This effect was not confined to pre-movement planning but also emerged during execution, indicating that the directional relationship between gaze and upcoming action remained sensitive to planning demands after movement had begun.

We next asked to what extent visual exploration aligns with the upcoming path across task phases and experimental manipulations, using two complementary measures of gaze-path alignment: \emph{spatial coverage}, which quantifies how much of the upcoming path was visually sampled, and \emph{temporal coherence}, which quantifies how closely the order of sampling matched the order of subsequent actions. Our results show that difficulty affected spatial coverage and temporal coherence in different and partly opposing, ways. Harder problems were associated with greater spatial coverage of the upcoming path, indicating that gaze sampled a larger portion of the route that participants subsequently executed. Before movement and during pauses, this increase was largely explained by the greater number of fixations elicited by difficult problems.

However, additional sampling did not increase temporal coherence. Instead, higher fixation counts were associated with lower coherence, indicating that broader sampling came at the cost of a weaker correspondence between fixation order and subsequent action order. Imposing a temporal delay similarly increased spatial coverage but not temporal coherence before movement. Together, these findings suggest that greater planning demands broadened visual exploration and spatial coverage of the upcoming path, particularly when a delay was imposed. However, the additional exploration elicited by difficulty was less sequential, revealing a tension between the spatial coverage and temporal ordering of plan-related information in gaze. These results raise the speculative possibility that, when planning demands increase, participants prioritize representing a broader portion of the upcoming path over encoding it in the order in which it will be executed. Future studies will be needed to test this possibility.

At the individual level, we found little evidence for discrete behavioural types. Instead, our analysis identified a continuous planning--revision dimension: participants with longer pre-movement inspection times and greater coverage of the upcoming path tended to produce fewer pauses and backtracks during execution. Advance planning and online revision therefore appeared not to characterize distinct groups, but rather to represent continuously varying tendencies in how participants distributed planning between pre-movement inspection and online revision. This interpretation should nevertheless remain tentative, given the limited sample size and modest between-participant stability of most measures.

Taken together, our findings show that the level of planning and its alignment with subsequent path execution, as revealed by visual exploration, are shaped by task difficulty and temporal constraints in different ways. Greater difficulty broadened visual exploration and spatial coverage of the upcoming path and increased repeated sampling, but resulted in less consistently sequential exploration. The presence or absence of a delay, in contrast, primarily shifted the distribution of planning across task phases. When participants could act immediately, they showed less visual exploration before movement but greater online revision, reflected in more pauses and backtracks. When a delay was imposed, more visual exploration occurred before movement, with fewer subsequent pauses and backtracks during execution. Thus, difficulty shaped the extent and organization of plan-related information in gaze, whereas delay shifted the balance between pre-movement plan construction and online revision. These findings characterize planning as an adaptive process whose visual and behavioural expression is flexibly distributed between preparation and execution according to task demands and temporal constraints.

This study has several limitations that could be addressed in future research. First, it will be important to assess the extent to which the present results generalize to planning problems with different characteristics, including richer and more ecologically realistic tasks requiring sophisticated visual strategies \citep{maselli2023beyond} and problems for which a problem space has to be constructed de novo \citep{liu2026beyond}. Evaluating the effects of misleadingness and  delay across a broader range of planning tasks may provide a better characterization of planning strategies and recurring sources of difficulty, while informing future computational models of planning. A related direction would be to compare difficulty induced by misleadingness with difficulty arising from structural properties of the problems, such as the size of the grid or the number of rewards, which were held constant in the present study. A further limitation concerns the planar, grid-like structure and restricted size of the maps used in this study. Dense connectivity may encourage fine-grained planning, but it also makes the tasks visually cluttered, complicating attempts to infer the underlying planning processes from gaze. Future research could employ alternative task designs that reduce these constraints. Several methodological limitations are also noteworthy, particularly regarding the clustering algorithm employed in this study. The algorithm was designed to reflect specific assumptions about how plans might be structured in this task and how such structures would manifest in gaze behaviour. Its parameters were validated using synthetic data and empirical considerations, giving the approach a largely top-down character. This choice was motivated by the covert nature of planning, which precludes direct access to ground-truth plans and their associated gaze dynamics. Behaviour during execution provides limited ground-truth in this respect because gaze during execution may reflect online action guidance rather than the structure of the preceding plan. One promising direction for future work would be to evaluate multiple algorithms embodying comparable assumptions about the mapping between plans and gaze and to assess the robustness of the results across methods.

A related limitation concerns the level at which condition-dependent changes in gaze can be interpreted. In particular, although the  delay manipulation redistributed repeated inspection and behavioural revision between preparation and execution, this does not establish that participants adopted qualitatively different visual strategies. The observed differences may instead reflect a temporal redistribution of the same underlying planning operations, with more processing occurring before movement when preparation time was available and more being deferred to online control when participants could act immediately. Similarly, re-fixations, reverse-directed sequences, and sampling outside the subsequently executed path are compatible with verification, consolidation, or exploration of alternatives, but do not uniquely identify these processes. Distinguishing changes in strategy from changes in the timing or intensity of a common planning mechanism will require designs that directly contrast competing computational accounts.

Finally, future work could record participants' brain activity while they perform the task to shed light on the neural mechanisms underlying planning. Previous studies have investigated how plan elements are represented in the brains of humans, non-human primates, and rodents, across brain regions including the frontal cortex and hippocampus \citep{xie2022geometry, chen2024flexible, mushiake2006activity,el2024cellular,Pfeiffer2020HippocampalReplay, Wikenheiser2015ThetaSequences,balaguer2016neural,mattar2022planning,donnarumma2026inferential}. Understanding how neural dynamics unfold during complex problem solving, and how they relate to behavioural strategies, remains an important objective for future research.

\section*{Data and Scripts Availability}
Data and scripts used for the analysis of this work are stored at \url{https://osf.io/dvrbu/overview?view_only=692be9ee85b547babffa48f31d6a4c07}.

\section*{Author Declaration}
We have no competing interests to declare.

\section*{AI-Use disclosure}
The authors declare that AI was used for grammar, spell check and text revision in this work. Authors take full responsibility for the content.

\section*{Acknowledgments}
This research received funding from the European Research Council under the Grant Agreement No. 820213 (ThinkAhead), the Italian National Recovery and Resilience Plan (NRRP), M4C2, funded by the European Union – NextGenerationEU (Project IR0000011, CUP B51E22000150006, “EBRAINS-Italy”; Project PE0000013, “FAIR”; Project PE0000006, “MNESYS”), and the Ministry of University and Research, PRIN PNRR P20224FESY and PRIN 20229Z7M8N. The GEFORCE Quadro RTX6000 and Titan GPU cards used for this research were donated by the NVIDIA Corporation.

\clearpage

\bibliographystyle{apalike}
\bibliography{references}
\clearpage

\FloatBarrier
\section*{Supplementary Materials}
\label{sec:supplementary_materials}

\setcounter{figure}{0}
\setcounter{table}{0}

\renewcommand{\thefigure}{S\arabic{figure}}
\renewcommand{\thetable}{S\arabic{table}}

\renewcommand{\figureautorefname}{Figure}
\renewcommand{\tableautorefname}{Table}

\subsection*{Structural properties of the maps}

\begin{figure}[ht!]
    \centering
    \includegraphics[width=1\linewidth]{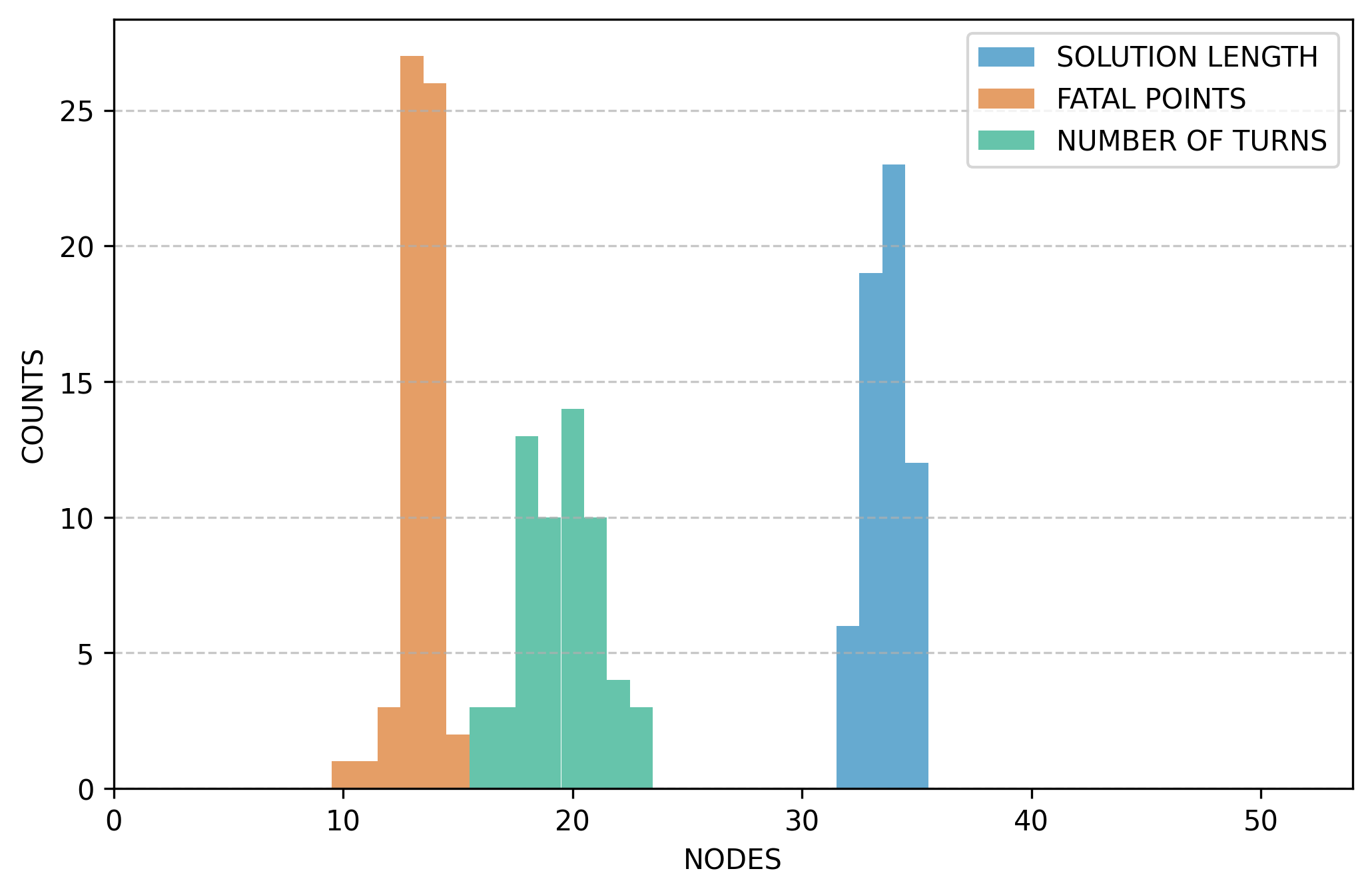}
    \caption{Distribution of key structural properties of the map. Solution length denotes the number of nodes in the shortest path; fatal points are nodes along the shortest path from which any deviation renders the task unsolvable; number of turns is computed along the shortest solution. All measures have an upper bound equal to the total number of nodes in the map (54).}
    \label{fig:struct_prop}
\end{figure}
\clearpage

\subsection*{The clustering algorithm} To infer the latent structure of the fixations while satisfying these requirements, we employed the generative-embedded Chinese Restaurant Process (geCRP) \citep{Maisto2016}. The geCRP defines a non-parametric clustering prior over fixations. This variant generalizes the distance-based Chinese Restaurant Process by replacing distance with a model-based measure of similarity between objects. To construct this prior, one must: (1) define a probability measure over individual objects or pairs of objects; and (2) define a kernel function that maps their probabilistic representations to similarity scores. Once the prior has been defined, a complete inference algorithm can be used to sample the cluster assignments of the fixations. 
\textbf{A structure-sensitive algorithmic probability for pairs of states}. Following \cite{Maisto2016}, we adopted the Algorithmic Probability (AlP) framework and assumed that the joint probability that two states in the graph are generated together is determined by the number of simple paths connecting them, with each path weighted by a factor that penalizes its length: 

\begin{equation}
P(s_i,s_l) \propto \sum_k \sum_j 2^{-\left(|c_i(s_k,\pi_j)| + |c_l(s_k,\pi_j)|\right)} = \sum_j 2^{-|c_i(s_l,\pi_j)|}. 
\end{equation}

To incorporate a basic constraint imposed by the task, we excluded all paths in which the starting node appeared at an intermediate position rather than at one of the endpoints. 
\textbf{Kernel definition.} The kernel maps the probabilistic representation of states or pairs of states to a similarity score. We considered a function comprising the product of two terms. The first term was: 

\begin{equation}
K_{\mathrm{AP}}(i,j) = P\bigl(s(i),s(j)\bigr)^2,
\end{equation}

where $s(i)$ and $s(j)$ are the $i$-th and $j$-th nodes fixated in the sequence, respectively. Pairs of nodes with a higher joint algorithmic probability therefore receive a higher similarity score. The second term was: 

\begin{equation}
K_{\mathrm{temporal}}(i,j) \propto \exp\left(-\frac{|i-j|}{\tau}\right). 
\end{equation}

This term assigns greater similarity to fixations that occur close together in the sequence. The resulting kernel, obtained by multiplying the two terms, is symmetric:

\begin{equation}
K(i,j) = K_{\mathrm{AP}}(i,j) K_{\mathrm{temporal}}(i,j).
\end{equation}

\textbf{Collapsed Gibbs-sampling algorithm.} Inference follows the standard dd-CRP framework, adapted to use the model-based similarity kernel rather than a distance function. Each state $s(i)$ is associated with a latent variable $c_i$ indicating the state to which it is linked. Its prior distribution is given by:

\begin{equation}
p(c_i=j \mid K,\alpha) \propto \begin{cases} \alpha, & \text{if } j=i,\\ K(i,j), & \text{if } j\neq i. \end{cases} \end{equation}

As in the standard CRP, the concentration parameter $\alpha$ controls the probability that an element is assigned to itself. We scaled $\alpha$ by sequence length to prevent cluster assignments from being systematically influenced by this extensive property of the sequence. The partition into clusters is induced by the complete set of link assignments: each cluster consists of a connected component of states. The clustering problem can therefore be formulated as the inference of the posterior distribution over link assignments given the observed data. Although direct evaluation is intractable because it requires estimating the model evidence, a collapsed Gibbs sampler can be used to update individual assignments from their analytically tractable conditional distributions while holding all other assignments fixed. Specifically, at each step, we remove one link and sample a new assignment conditional on all remaining links:

\begin{equation}
p\!\left(c_i^{\mathrm{new}} \mid c_{-i}, s_{1:N}, K, \alpha\right) \propto \begin{cases} \alpha, & \text{if } c_i^{\mathrm{new}} = i,\\[4pt] K(i,j), & \text{if } c_i^{\mathrm{new}} = j \text{ does not merge two clusters},\\[4pt] K(i,j)\,\Gamma(s,z,G_0), & \text{if } c_i^{\mathrm{new}} = j \text{ merges clusters } k \text{ and } l. \end{cases} \end{equation}

Here, $\Gamma$ is the ratio between the likelihood of the proposed cluster configuration in which clusters $k$ and $l$ are merged and the likelihood of the configuration in which they remain separate: \begin{equation} \Gamma(s,z,G_0) = \frac{ p\!\left( s_{z(c_{-i})=k \,\cup\, z(c_{-i})=l} \mid G_0 \right) }{ p\!\left(s_{z(c_{-i})=k}\mid G_0\right) p\!\left(s_{z(c_{-i})=l}\mid G_0\right) }. \end{equation} The marginal likelihood of the observations assigned to cluster $k$ is given by:

\begin{equation}
p\!\left( s_{z(c_{1:N})=k} \mid z(c_{1:N}),G_0 \right) = \sum_{j\neq i} \left[ \mathbb{I}\!\left(i\in s_{z(c_{1:N})=k}\right) \frac{ \sum_{j\neq i}K(i,j)+1 }{ n_k+N } \right]. 
\end{equation}

\subsection*{Inference and clustering of the co-occurrence matrix}

The (collapsed) Gibbs-sampling algorithm provides a putative assignment of fixations to clusters. By sampling multiple clustering configurations after the algorithm has reached equilibrium, we can construct a co-occurrence matrix for the states. For our simulations, we ran the algorithm for a total of $10^4$ steps across 50 replicas of the inference state, each initialized with a randomized configuration. We used the preceding equation to identify the clustering with the highest likelihood in each replica. For each fixation sequence of length $n$, the co-occurrence matrix was transformed into a cosine similarity matrix through row-wise normalization. A distance matrix was then defined as $\text{distance} = 1 - \text{similarity}$. The $i$-th row of this matrix represents the distances between the $i$-th fixation and all other elements in the sequence, as induced by the inference algorithm and its underlying model. Agglomerative hierarchical clustering was performed using Ward’s minimum variance criterion, which identifies $K$ clusters by minimizing the total within-cluster variance. Although Ward’s method is formally defined for Euclidean distances, an empirical evaluation of the resulting clusters supported its suitability for this setting. The algorithm requires the number of clusters $K$ to be specified in advance. To determine an appropriate value, we evaluated values of $K$ from $1$ to $n$ and computed the clustering inertia for each value. Inertia was defined as the sum of the squared deviations of each point from its cluster mean, aggregated across clusters. By construction, inertia decreases as $K$ increases, with progressively diminishing gains. The elbow point---the value of $K$ beyond which additional clusters provide negligible improvement---was identified as the first $K$ for which inertia fell below a fixed fraction (threshold ratio $= 0.1$) of the maximum inertia observed for the sequence (i.e., at $K = 1$, when all fixations belong to a single cluster). This threshold was selected empirically based on experiments with both synthetic and real data and proved robust across varying sequence lengths. In some sequences, fixations were spatially distant from all other elements, resulting in near-zero similarity values. To detect such outliers, we computed the maximum pairwise similarity for each node. If this value was below an empirical threshold (maximum similarity threshold $=10^{-5}$), the node was reassigned to a separate singleton cluster. This dual-threshold strategy---combining elbow-based model selection with explicit outlier isolation---balanced parsimony with clustering quality, preventing both over-segmentation and the inappropriate grouping of temporally or behaviourally distinct fixation patterns.
\clearpage

\subsection*{Validation of the Clustering Pipeline}
 
The geCRP and elbow-clustering pipeline used to segment participants' fixation sequences into planning sub-sequences (see \nameref{sec:methods}) has no external ground truth on real data: we cannot know, independently of the algorithm, which fixations truly belong together. To establish that the pipeline is capable of recovering meaningful structure in gaze patterns, we validated it on synthetic fixation sequences with specific spatio-temporal configurations, corresponding to a set of predetermined cases. Sequences were generated directly on a set of graphs of the real experimental maps and processed through the identical pipeline used on the experimental data.
 
The cases we considered for the validation belong to what \textit{a priori} can be interpreted as plausible patterns of fixations in a task like ThinkAhead and allowed for the assessment of two different properties: \emph{sensitivity}, whether the pipeline recovers structure that is genuinely present, and \emph{specificity}, whether it refrains from inventing structure that is not. For structured cases, recovered labels were compared to ground truth using the Adjusted Rand Index (ARI; \citealp{hubert1985comparing}). The ARI measures how consistently pairs of fixations are assigned to the same or different clusters in the recovered and ground-truth partitions, while correcting for agreement expected by chance. An ARI of 1 indicates perfect recovery, 0 indicates chance-level agreement, and negative values indicate worse-than-chance agreement. For null cases, recovery was scored as the false-positive rate: the fraction of instances in which the pipeline reported more than one cluster. We next describe the cases considered in the validation.

\textbf{\textit{Close group}}. A single, spatially compact set of fixations around one map location, with one ground-truth cluster. The desired outcome would be that fixations that are close together in the graph should not be split into spurious sub-groups.

\textbf{\textit{Distinct groups}}. Several spatially well-separated, compact clusters, generalizing the previous case to more than one concurrent group. Because participants may consider more than one candidate path or reward location at a time, the pipeline must recover an arbitrary, \textit{a priori} unknown number of simultaneous clusters, not merely detect the presence of one.

\textbf{\textit{Walk with interruptions}}. A single contiguous exploratory walk, mimicking inspection of one candidate path, interspersed with brief glances to one far-away region; two ground-truth clusters result (the walk and the interruption). This reproduces a common real pattern, a participant briefly checking an alternative location while primarily tracing one path, and tests whether the pipeline separates temporally interleaved but spatially distinct gaze.

\textbf{\textit{Bifurcation}}. A shared approach segment leading to a fork, followed by two alternative continuations, only one of which continues the approach. This emulates a genuine planning decision point, where gaze considers two viable onward routes from a common location. It is a stricter test than distinct groups because the two ground-truth clusters share a proximal segment and are not globally far apart on the map.

\textbf{\textit{Long walk}}. A single long self-avoiding walk with no embedded substructure. The correct outcome is exactly one cluster; the pipeline is scored on how often it incorrectly reports more than one, which would indicate that continuous exploration of a single path is being spuriously fragmented. Note, however, that if the clustering pipeline splits the long walk into multiple clusters, the subsequent recombination procedure may still recover the complete structure by identifying the best-fitting sequence.

\textbf{\textit{Random}}. Fixations placed at random, unstructured locations, with no spatial or temporal coherence at all. This complements the long-walk control by probing a different failure mode: whether scattered, plan-free gaze is nonetheless assembled into spurious clusters.

\begin{figure}
    \centering
    \includegraphics[width=\linewidth]{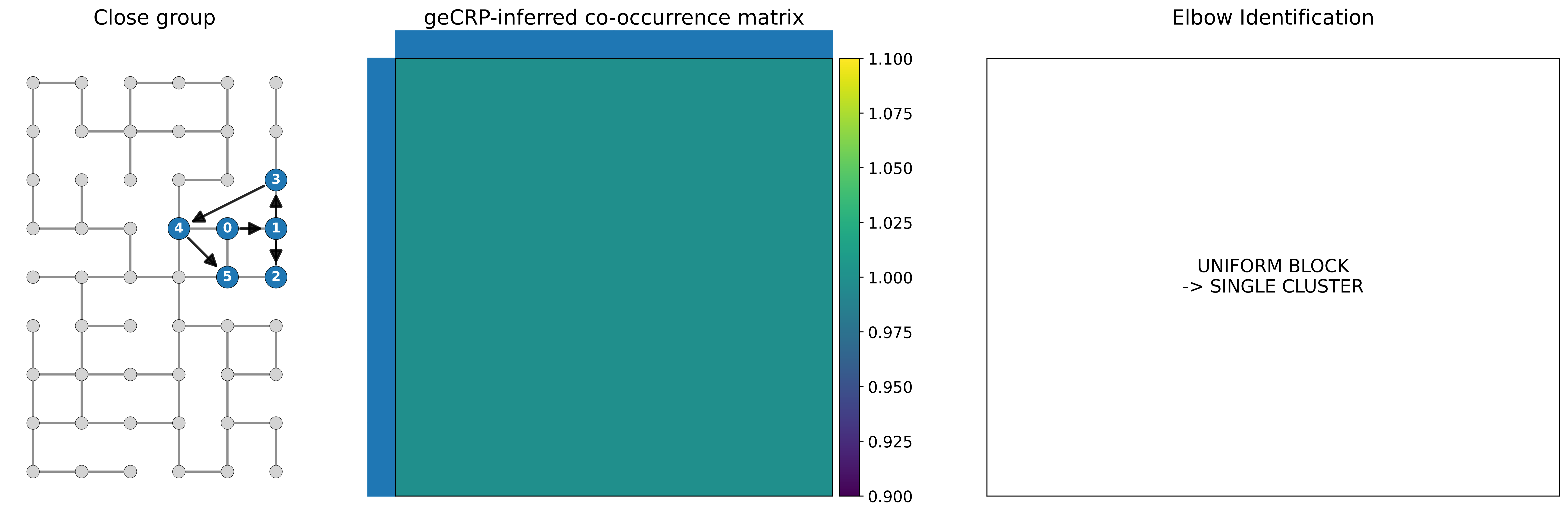}
    \includegraphics[width=\linewidth]{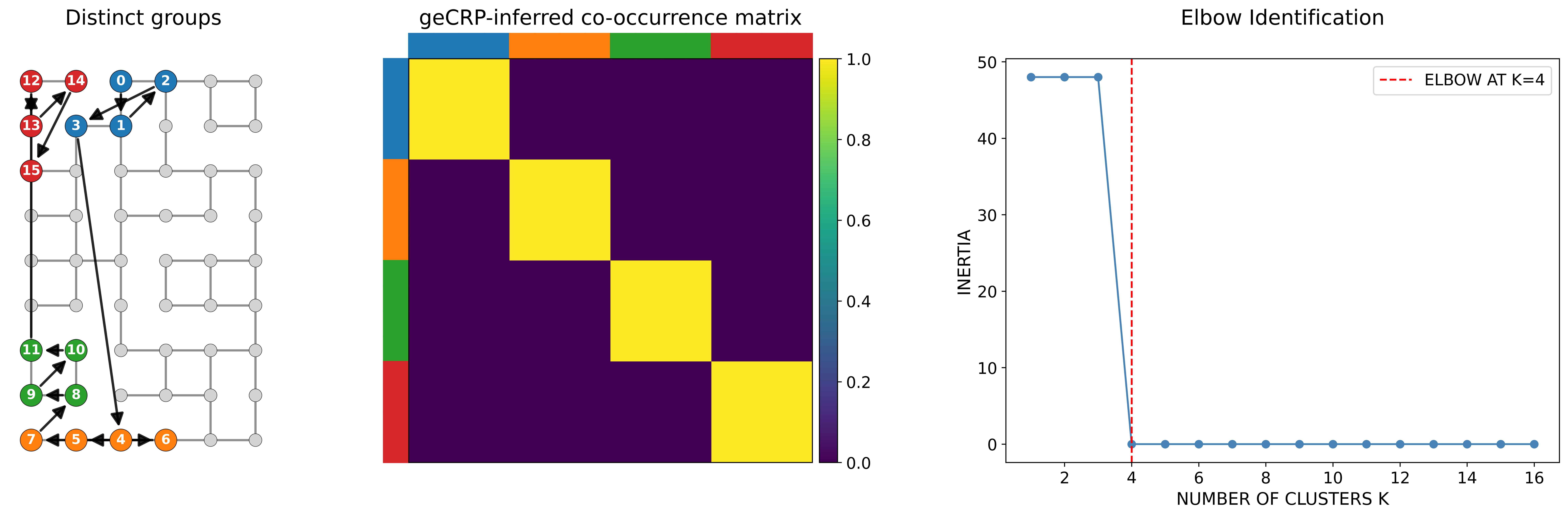}
    \includegraphics[width=\linewidth]{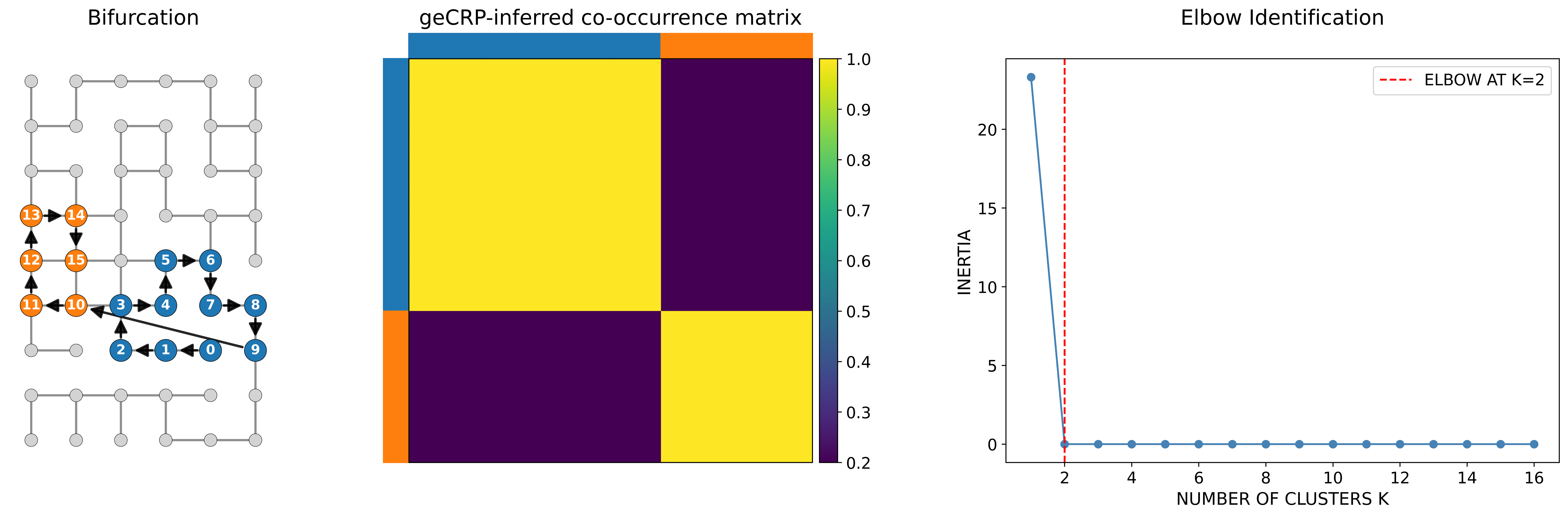}
    \caption{Validation of the clustering pipeline on synthetic gaze sequences. Each row shows one validation case processed by the full geCRP + elbow-clustering pipeline on the graph of a real experimental map; from top to bottom: close group,
    distinct groups, bifurcation. Left: the synthetic fixation sequence overlaid on the map graph (grey nodes and edges). Each colored circle represents one fixation; the number inside it is the position of that fixation in the sequence (0 = first fixation), and the arrows connect consecutive fixations, pointing from each fixation to the next one in time. Circle color indicates the ground-truth cluster. Middle: the geCRP co-occurrence matrix inferred from the sequence, with fixations in temporal order and the pipeline's recovered clusters indicated by the color bands along the axes. Right: the elbow criterion used to select the number of
    clusters, i.e. hierarchical-clustering inertia as a function of the candidate number of clusters $k$ (dashed line: selected elbow); when the co-occurrence matrix is a single uniform block, the elbow step is skipped and one cluster is returned. Each case is shown at the operating point where the pipeline performs best in the parameter sweeps (close group: size 6; distinct groups: $k=4$ groups of 4; bifurcation: branch length 6).}
    \label{fig:clustering_validation1}
\end{figure}

\begin{figure}
    
    \includegraphics[width=\linewidth]{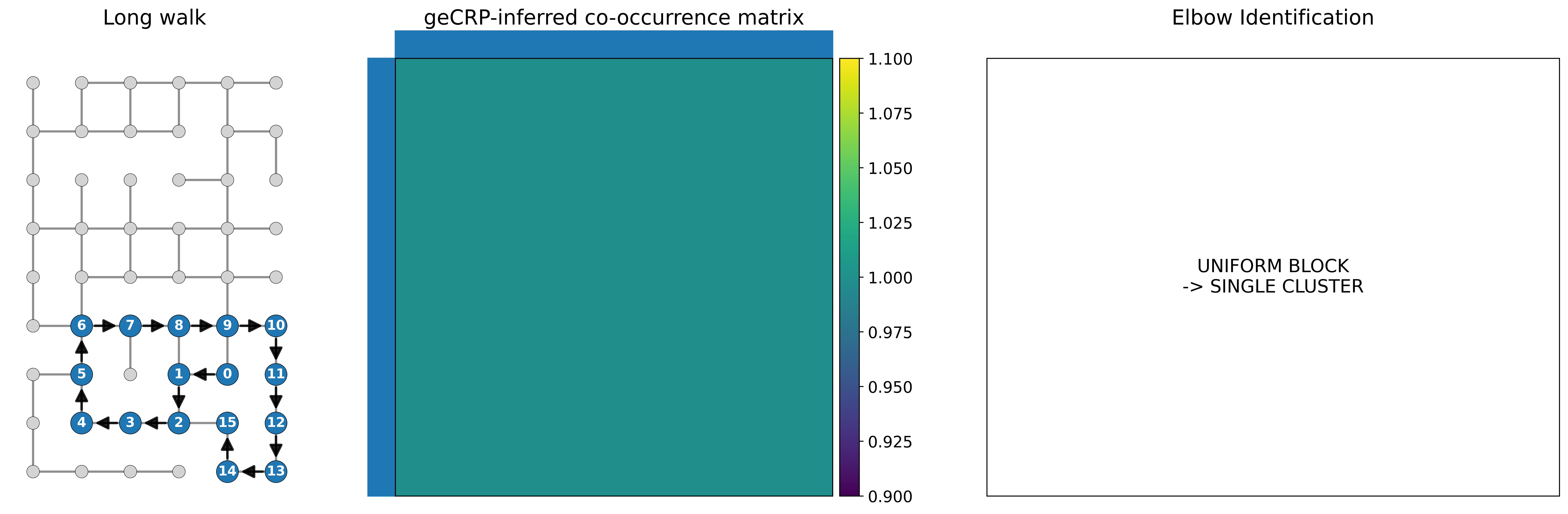}
    \includegraphics[width=\linewidth]{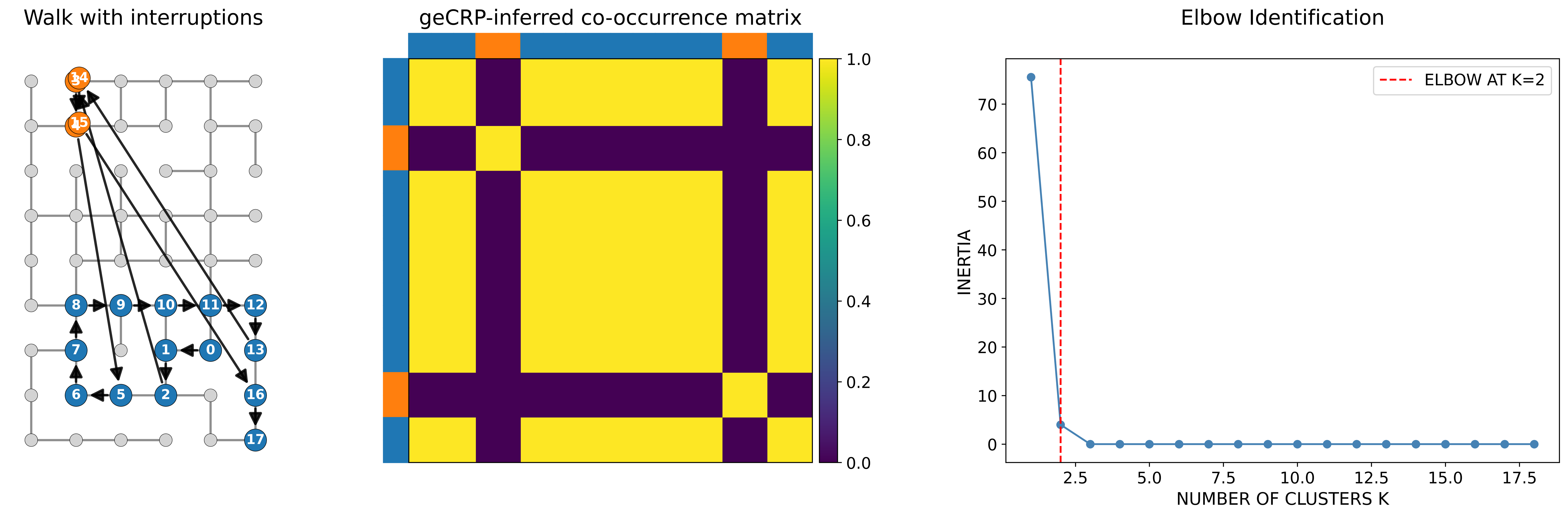}
    \includegraphics[width=\linewidth]{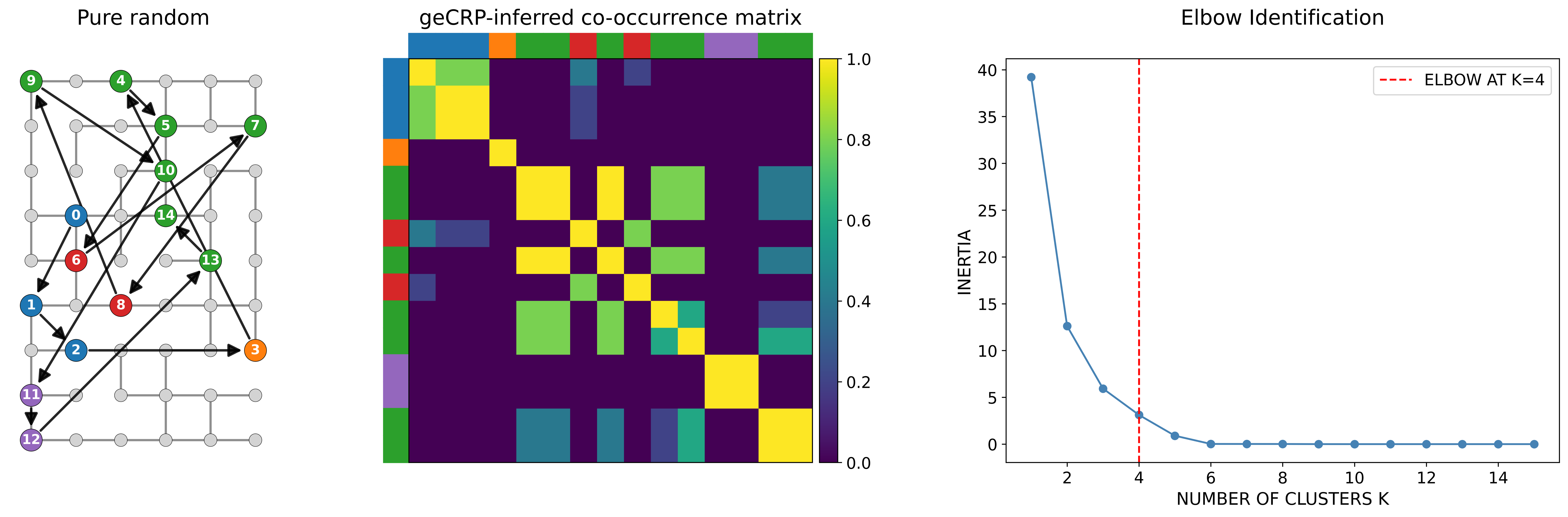}
    \caption{Validation of the clustering pipeline on synthetic gaze sequences. Each row shows one validation case processed by the full geCRP + elbow-clustering pipeline on the graph of a real experimental map; from top to bottom: long walk, walk with interruptions, and random. Each case is shown at the operating point where the pipeline performs best in the parameter sweeps (long walk: length 16; walk with interruptions: length 14).}
    \label{fig:clustering_validation2}
\end{figure}

To assess robustness rather than only best-case recovery, each structured case was perturbed in two orthogonal directions: a temporal reshuffling, and an increase in cluster size.
For the controlled temporal reshuffle, within each ground-truth subsequence, a fraction $r$ of fixations was reordered while their node identities and cluster membership were kept fixed. $r=0$ reproduces the unperturbed case exactly; $r=1$ fully randomizes the order within each subsequence. This isolates how much of the pipeline's recovery relies on temporal proximity as opposed to spatial co-occurrence and characterizes recovery as a function of reshuffle rate rather than at a single operating point.
For each structured case, the parameter defining the scale of its ground-truth structure (the size of the single group, the number of distinct groups, the length of the interrupted walk, the length of the bifurcation branches) was varied systematically rather than fixed at one arbitrary value. Recovery can depend sharply on this scale: for instance, two branches sharing a fork are expected to be easier to tell apart the longer they are before merging back near the fork, so a single branch length could give a misleadingly optimistic or pessimistic picture of pipeline performance. 
Visual representations of how these cases are processed are provided in \autoref{fig:clustering_validation1} and \autoref{fig:clustering_validation2}.

\FloatBarrier
\FloatBarrier

We now present the results of the robustness analysis across the different synthetic cases and parameter settings. For each synthetic case, we generated 30 instances (6 maps $\times$ 5 random seeds); each instance was processed by the full pipeline, running the geCRP with concentration parameter $\alpha = 1.5/n$ ($n$ = number of fixations in the sequence), 500 Gibbs iterations and 5 replicas, whose partitions were averaged into a single co-occurrence matrix before elbow clustering, and scored by a single Adjusted Rand Index against the ground-truth labels. Tables report the mean $\pm$ SD across instances.
Reshuffle conditions used 5 independent permutations per instance (150 instances per cell); the unperturbed condition is deterministic and therefore run once per instance (30 instances).

\textit{\textbf{Close group}}. Recovery degraded sharply as the single group grew larger, in both the unperturbed case and under reshuffling (\autoref{tab:closegroup}); larger single clusters are increasingly prone to being spuriously split.

\begin{table}[htbp]
\centering
\begin{tabular}{lcccc}
\hline
Size & rate 0 & rate 0.1 & rate 0.25 & rate 0.5 \\
\hline
6 & 0.97 & 0.97 & 0.96 & 0.97 \\
9 & 0.83 & 0.83 & 0.82 & 0.75 \\
12 & 0.77 & 0.80 & 0.61 & 0.46 \\
\hline
\end{tabular}
\caption{Close group: Adjusted Rand Index as a function of group size and reshuffle rate.}
\label{tab:closegroup}
\end{table}

\textit{\textbf{Distinct groups}}. Recovery degraded moderately as the number of concurrent groups $k$ increased, and became markedly more fragile to reshuffling at larger $k$ (\autoref{tab:kswep}).

\begin{table}[htbp]
\centering
\begin{tabular}{lcccc}
\hline
$k$ & rate 0 & rate 0.1 & rate 0.25 & rate 0.5 \\
\hline
3 & 0.94 & 0.96 & 0.90 & 0.87 \\
4 & 0.96 & 0.89 & 0.82 & 0.72 \\
5 & 0.85 & 0.83 & 0.76 & 0.61 \\
\hline
\end{tabular}
\caption{Distinct groups: Adjusted Rand Index as a function of the number of groups $k$ and reshuffle rate (group size fixed at 4).}
\label{tab:kswep}
\end{table}

\textit{\textbf{Bifurcation}}. Bifurcation transitioned sharply from failure to success as branch length increased (\autoref{tab:bifurcation}): ARI rose from $0.07$ at length 4 to $\approx 0.90$ at length 6, with no further gain at length 8. Even once recoverable, bifurcation remained the structured case most fragile to reshuffling, degrading by roughly half from rate 0 to rate 0.5 at both lengths 6 and 8.

\begin{table}[htbp]
\centering
\begin{tabular}{lcccc}
\hline
Branch length & rate 0 & rate 0.1 & rate 0.25 & rate 0.5 \\
\hline
4 & 0.07 & 0.09 & 0.19 & 0.16 \\
6 & 0.90 & 0.88 & 0.65 & 0.37 \\
8 & 0.91 & 0.85 & 0.71 & 0.42 \\
\hline
\end{tabular}
\caption{Bifurcation: Adjusted Rand Index as a function of branch length and reshuffle rate.}
\label{tab:bifurcation}
\end{table}

\textit{\textbf{Long walk}}. The long-walk null case correctly yielded a single cluster at walk lengths 8 and 16, but this broke down at length 24, where more than half of the instances were incorrectly split into multiple clusters (\autoref{tab:longsaw}). 

\begin{table}[htbp]
\centering
\begin{tabular}{lccc}
\hline
Length & False-positive rate & Mean clusters \\
\hline
8 & 0.00 & 1.00 \\
16 & 0.00 & 1.00 \\
24 & 0.57 & 1.60 \\
\hline
\end{tabular}
\caption{Long walk: false-positive rate (fraction of instances reporting more than one cluster) and mean recovered cluster count, as a function of walk length.}
\label{tab:longsaw}
\end{table}
\FloatBarrier
\textit{\textbf{Random}}. The purely random case was reliably over-segmented at all tested settings (false-positive rate $1.0$, 4--7 clusters; not shown).

\clearpage

\subsection*{Misleadingness increases planning demand, whereas  delay reduces online revision}

\begin{table}[ht]
\centering
\small
\begin{tabular*}{\textwidth}{@{\extracolsep{\fill}}llrrrrl}
\hline
\multicolumn{7}{l}{\footnotesize\textit{$a$: IV $\rightarrow$ Mediator; $c$: total effect (IV $\rightarrow$ DV); $c'$: direct effect; $a \times b$: indirect effect; $b$: Mediator $\rightarrow$ DV}} \\
\hline
IV & Path & Estimate & Std.Err & p-value & 95\% CI & \\
\hline\\
Difficulty & $a$ & $0.3870$ & $0.0550$ & $<10^{-9}$ & --- & *** \\
 & $c$ & $-2.672$ & $0.2489$ & $<10^{-9}$ & --- & *** \\
 & $c'$ & $-2.610$ & $0.2501$ & $<10^{-9}$ & --- & *** \\
 & $a \times b$ & $-0.0750$ (2.8) & $0.0315$ & $0.0171$ & [$-0.1560$, $-0.0138$] & * \\
\\
 delay & $a$ & $-0.7201$ & $0.0563$ & $<10^{-9}$ & --- & *** \\
 & $c$ & $-0.2642$ & $0.3082$ & $0.3914$ & --- &  \\
 & $c'$ & $-0.4004$ & $0.3140$ & $0.2023$ & --- &  \\
 & $a \times b$ & $0.1396$ (-52.8) & $0.0561$ & $0.0129$ & [$0.0319$, $0.2605$] & * \\
\\
Difficulty $\times$  delay & $a$ & $-0.1677$ & $0.0795$ & $0.0351$ & --- & * \\
 & $c$ & $0.2064$ & $0.3428$ & $0.5471$ & --- &  \\
 & $c'$ & $0.1723$ & $0.3439$ & $0.6164$ & --- &  \\
 & $a \times b$ & $0.0325$ (15.8) & $0.0201$ & $0.1050$ & [$0.00254$, $0.0857$] &  \\
\\
Delay-adj.\ view time & $b$ & $-0.1939$ & $0.0765$ & $0.0112$ & --- & * \\
Level (covariate) & direct & $0.5441$ & $0.1365$ & $6.7\times10^{-5}$ & --- & *** \\
\hline
\end{tabular*}
\caption{Mediation analysis: Delay-adjusted View Time mediating the effects of Difficulty,  delay, and their interaction on Success. Path $a$: LMM, random intercept per participant; paths $c$, $c'$, $b$: logistic GLMM with random intercept per participant (estimates in log-odds); $a \times b$: indirect effect (Sobel test for SE/p-value; bootstrap 95\% CI from cluster resampling); proportion mediated ($a \times b / c$) shown in parentheses. The p-value significance code: $*** < 10^{-3}$, $10^{-3} < ** < 10^{-2}$, $10^{-2} < * < 0.05$, $0.05 < \text{.} < 0.1$, $> 0.1$.}
\label{tab:mediation_success}
\end{table}

\begin{table}[ht]
\centering
\small
\begin{tabular*}{\textwidth}{@{\extracolsep{\fill}}llrrrrl}
\hline
\multicolumn{7}{l}{\footnotesize\textit{$a$: IV $\rightarrow$ Mediator; $c$: total effect (IV $\rightarrow$ DV); $c'$: direct effect; $a \times b$: indirect effect; $b$: Mediator $\rightarrow$ DV}} \\
\hline
IV & Path & Estimate & Std.Err & p-value & 95\% CI & \\
\hline\\
Difficulty & $a$ & $0.3870$ & $0.0550$ & $<10^{-9}$ & --- & *** \\
 & $c$ & $1.316$ & $0.0918$ & $<10^{-9}$ & --- & *** \\
 & $c'$ & $1.350$ & $0.0928$ & $<10^{-9}$ & --- & *** \\
 & $a \times b$ & $-0.0421$ (-3.2) & $0.0176$ & $0.0171$ & [$-0.0644$, $-0.0167$] & * \\
\\
Delay & $a$ & $-0.7201$ & $0.0563$ & $<10^{-9}$ & --- & *** \\
 & $c$ & $-0.3113$ & $0.1066$ & $0.0035$ & --- & ** \\
 & $c'$ & $-0.3755$ & $0.1093$ & $5.9\times10^{-4}$ & --- & *** \\
 & $a \times b$ & $0.0783$ (-25.1) & $0.0315$ & $0.0129$ & [$0.0337$, $0.1187$] & * \\
\\
Difficulty $\times$ Delay & $a$ & $-0.1677$ & $0.0795$ & $0.0351$ & --- & * \\
 & $c$ & $-0.0620$ & $0.1370$ & $0.6508$ & --- &  \\
 & $c'$ & $-0.0797$ & $0.1368$ & $0.5599$ & --- &  \\
 & $a \times b$ & $0.0182$ (-29.4) & $0.0112$ & $0.1050$ & [$0.00217$, $0.0340$] &  \\
\\
Delay-adjusted view time & $b$ & $-0.1087$ & $0.0429$ & $0.0113$ & --- & * \\
Block (covariate) & direct & $-0.0941$ & $0.0670$ & $0.1602$ & --- &  \\
\hline
\end{tabular*}
\caption{Mediation analysis: Delay-adjusted View Time mediating the effects of Misleadingness,  delay, and their interaction on Number of Backtracks. Path $a$: LMM, random intercept per participant; paths $c$, $c'$, $b$: Poisson GLMM with random intercept per participant (estimates in log-counts); $a \times b$: indirect effect (Sobel test for SE/p-value; bootstrap 95\% CI from cluster resampling); proportion mediated ($a \times b / c$) shown in parentheses. The p-value significance code: $*** < 10^{-3}$, $10^{-3} < ** < 10^{-2}$, $10^{-2} < * < 0.05$, $0.05 < \text{.} < 0.1$, $> 0.1$.}
\label{tab:mediation_num_backtrack}
\end{table}

\begin{table}[ht]
\centering
\begin{tabular*}{\textwidth}{@{\extracolsep{\fill}}lrrrrl}
\hline
Variables Predicting\\ Backtrack Rate (per time) & Estimate & Std.Err & t-statistic & p-value & \\
\hline\\
Intercept & $4.72\times10^{-5}$ & $5.15\times10^{-6}$ & $9.159$ & $<10^{-9}$ & *** \\
Block & $9.61\times10^{-7}$ & $3.24\times10^{-6}$ & $0.2965$ & $0.7669$ &  \\
Difficulty & $5.59\times10^{-5}$ & $3.29\times10^{-6}$ & $16.966$ & $<10^{-9}$ & *** \\
 delay & $-1.33\times10^{-5}$ & $3.63\times10^{-6}$ & $-3.654$ & $2.7\times10^{-4}$ & *** \\
Delay-adjusted view time & $-2.53\times10^{-6}$ & $2.03\times10^{-6}$ & $-1.242$ & $0.2143$ &  \\
\hline
\end{tabular*}
\caption{Backtrack Rate (per time) -- LMM. The p-value significance code: $*** < 10^{-3}$, $10^{-3} < ** < 10^{-2}$, $10^{-2} < * < 0.05$, $0.05 < \text{.} < 0.1$, $> 0.1$.}
\label{tab:num_backtrack_per_time}
\end{table}

\begin{table}[ht]
\centering
\begin{tabular*}{\textwidth}{@{\extracolsep{\fill}}lrrrrl}
\hline
Variables Predicting\\ Pause Rate (per time) & Estimate & Std.Err & t-statistic & p-value & \\
\hline\\
Intercept & $1.20\times10^{-4}$ & $9.06\times10^{-6}$ & $13.270$ & $<10^{-9}$ & *** \\
Block & $-1.95\times10^{-5}$ & $4.40\times10^{-6}$ & $-4.437$ & $9.8\times10^{-6}$ & *** \\
Difficulty & $6.10\times10^{-5}$ & $6.18\times10^{-6}$ & $9.866$ & $<10^{-9}$ & *** \\
 delay & $-3.16\times10^{-5}$ & $6.55\times10^{-6}$ & $-4.823$ & $1.6\times10^{-6}$ & *** \\
Delay-adjusted view time & $4.03\times10^{-6}$ & $2.79\times10^{-6}$ & $1.440$ & $0.1500$ &  \\
Difficulty $\times$  delay & $1.73\times10^{-5}$ & $8.80\times10^{-6}$ & $1.964$ & $0.0497$ & * \\
\hline
\end{tabular*}
\caption{Pause Rate (per time) -- LMM. The p-value significance code:  $*** < 10^{-3}$, $10^{-3} < ** < 10^{-2}$, $10^{-2} < * < 0.05$, $0.05 < \text{.} < 0.1$, $> 0.1$.}
\label{tab:num_pauses_per_time}
\end{table}
\clearpage

\subsection*{Visual exploration is shaped by task phase and difficulty}

\begin{table}[ht]
\centering
\begin{tabular*}{\textwidth}{@{\extracolsep{\fill}}lrrrrl}
\hline
START & Estimate & Std.Err & Wald z & p-value & \\
\hline\\
Intercept & $-2.1587$ & $0.0518$ & $-41.7143$ & $<10^{-9}$ & *** \\
Pause & $-2.2666$ & $0.1537$ & $-14.7434$ & $<10^{-9}$ & *** \\
Exec & $-4.0888$ & $0.1966$ & $-20.7937$ & $<10^{-9}$ & *** \\
Hard & $0.1064$ & $0.0511$ & $2.0813$ & $0.0374$ & * \\
 delay & $-0.1968$ & $0.0468$ & $-4.2105$ & $2.5\times10^{-5}$ & *** \\
Pause:Hard & $1.1861$ & $0.1643$ & $7.2197$ & $<10^{-9}$ & *** \\
Exec:Hard & $0.8447$ & $0.2255$ & $3.7468$ & $1.8\times10^{-4}$ & *** \\
Pause: delay & $0.6276$ & $0.2167$ & $2.8965$ & $0.0038$ & ** \\
Exec: delay & $-0.0300$ & $0.3031$ & $-0.0991$ & $0.9211$ &  \\
Hard: delay & $-0.0395$ & $0.0609$ & $-0.6480$ & $0.5170$ &  \\
Pause:Hard: delay & $-0.5294$ & $0.2336$ & $-2.2668$ & $0.0234$ & * \\
Exec:Hard: delay & $0.1590$ & $0.3461$ & $0.4593$ & $0.6460$ &  \\
\hline
\end{tabular*}
\caption{GLMM Binomial -- START Fixation Proportions (ref: Before Movement). Difficulty ref: Easy;  delay ref: No- delay. The p-value significance code: $*** < 10^{-3}$, $10^{-3} < ** < 10^{-2}$, $10^{-2} < * < 0.05$, $0.05 < \text{.} < 0.1$, $> 0.1$.}
\label{tab:glmm_start}
\end{table}

\begin{table}[ht]
\centering
\begin{tabular*}{\textwidth}{@{\extracolsep{\fill}}lrrrrl}
\hline
REWARD & Estimate & Std.Err & Wald z & p-value & \\
\hline\\
Intercept & $-0.9537$ & $0.0423$ & $-22.5260$ & $<10^{-9}$ & *** \\
Pause & $0.0684$ & $0.0444$ & $1.5399$ & $0.1236$ &  \\
Exec & $0.0439$ & $0.0324$ & $1.3537$ & $0.1758$ &  \\
Hard & $0.0708$ & $0.0344$ & $2.0545$ & $0.0399$ & * \\
 delay & $0.0215$ & $0.0305$ & $0.7047$ & $0.4810$ &  \\
Pause:Hard & $-0.1579$ & $0.0542$ & $-2.9123$ & $0.0036$ & ** \\
Exec:Hard & $-0.0985$ & $0.0427$ & $-2.3061$ & $0.0211$ & * \\
Pause: delay & $-0.1214$ & $0.0653$ & $-1.8604$ & $0.0628$ & . \\
Exec: delay & $-0.0425$ & $0.0411$ & $-1.0337$ & $0.3013$ &  \\
Hard: delay & $-0.0403$ & $0.0403$ & $-1.0003$ & $0.3171$ &  \\
Pause:Hard: delay & $0.0387$ & $0.0783$ & $0.4947$ & $0.6208$ &  \\
Exec:Hard: delay & $0.0843$ & $0.0548$ & $1.5368$ & $0.1243$ &  \\
\hline
\end{tabular*}
\caption{GLMM Binomial -- REWARD Fixation Proportions (ref: Before Movement). Difficulty ref: Easy;  delay ref: No- delay. The p-value significance code: $*** < 10^{-3}$, $10^{-3} < ** < 10^{-2}$, $10^{-2} < * < 0.05$, $0.05 < \text{.} < 0.1$, $> 0.1$.}
\label{tab:glmm_reward}
\end{table}

\begin{table}[ht]
\centering
\begin{tabular*}{\textwidth}{@{\extracolsep{\fill}}lrrrrl}
\hline
OTHER & Estimate & Std.Err & Wald z & p-value & \\
\hline\\
Intercept & $0.4857$ & $0.0410$ & $11.8377$ & $<10^{-9}$ & *** \\
Pause & $0.3419$ & $0.0430$ & $7.9585$ & $<10^{-9}$ & *** \\
Exec & $0.4142$ & $0.0309$ & $13.4206$ & $<10^{-9}$ & *** \\
Hard & $-0.1022$ & $0.0319$ & $-3.2018$ & $0.0014$ & ** \\
 delay & $0.0451$ & $0.0283$ & $1.5963$ & $0.1104$ &  \\
Pause:Hard & $0.0430$ & $0.0520$ & $0.8275$ & $0.4079$ &  \\
Exec:Hard & $0.1169$ & $0.0407$ & $2.8746$ & $0.0040$ & ** \\
Pause: delay & $0.0193$ & $0.0634$ & $0.3049$ & $0.7604$ &  \\
Exec: delay & $-0.0225$ & $0.0395$ & $-0.5687$ & $0.5696$ &  \\
Hard: delay & $0.0529$ & $0.0374$ & $1.4124$ & $0.1578$ &  \\
Pause:Hard: delay & $0.0005$ & $0.0755$ & $0.0064$ & $0.9949$ &  \\
Exec:Hard: delay & $-0.0986$ & $0.0527$ & $-1.8725$ & $0.0611$ & . \\
\hline
\end{tabular*}
\caption{GLMM Binomial -- OTHER Fixation Proportions (ref: Before Movement). Difficulty ref: Easy;  delay ref: No- delay. The p-value significance code: $*** < 10^{-3}$, $10^{-3} < ** < 10^{-2}$, $10^{-2} < * < 0.05$, $0.05 < \text{.} < 0.1$, $> 0.1$.}
\label{tab:glmm_other}
\end{table}

\begin{table}[ht]
\centering
\begin{tabular*}{\textwidth}{@{\extracolsep{\fill}}llllrrrrl}
\hline
 delay\\ Cond. & Difficulty & Node type & Contrast & Odds ratio & SE & z-ratio & p-value  & sig.\\
\hline\\
ND & Easy & S & BM > P & 9.6467 & 1.4831 & 14.7434 & $<10^{-3}$ & *** \\
ND & Easy & S & P > E & 6.1855 & 1.5035 & 7.4967 & $<10^{-3}$ & *** \\
D & Easy & S & BM > P & 5.1502 & 0.7869 & 10.7281 & $<10^{-3}$ & *** \\
D & Easy & S & P > E & 11.9392 & 3.2775 & 9.0333 & $<10^{-3}$ & *** \\
ND & Hard & S & BM > P & 2.9462 & 0.1751 & 18.1774 & $<10^{-3}$ & *** \\
ND & Hard & S & P > E & 8.7020 & 1.0130 & 18.5851 & $<10^{-3}$ & *** \\
D & Hard & S & BM > P & 2.6708 & 0.1753 & 14.9705 & $<10^{-3}$ & *** \\
D & Hard & S & P > E & 8.4384 & 1.1631 & 15.4732 & $<10^{-3}$ & *** \\
ND & Easy & R & BM / P & 0.9339 & 0.0415 & -1.5399 & $3.7\times 10^{-1}$ &  \\
ND & Easy & R & BM / E & 0.9570 & 0.0311 & -1.3537 & $3.7\times 10^{-1}$ &  \\
ND & Easy & R & P / E & 1.0248 & 0.0414 & 0.6068 & $5.4\times 10^{-1}$ &  \\
D & Easy & R & BM / P & 1.0545 & 0.0505 & 1.1066 & $8.1\times 10^{-1}$ &  \\
D & Easy & R & BM / E & 0.9986 & 0.0253 & -0.0555 & $9.6\times 10^{-1}$ &  \\
D & Easy & R & P / E & 0.9470 & 0.0470 & -1.0970 & $8.1\times 10^{-1}$ &  \\
ND & Hard & R & BM / P & 1.0936 & 0.0347 & 2.8204 & $1.4\times 10^{-2}$ & * \\
ND & Hard & R & BM / E & 1.0561 & 0.0300 & 1.9251 & $1.1\times 10^{-1}$ &  \\
ND & Hard & R & P / E & 0.9658 & 0.0268 & -1.2559 & $2.1\times 10^{-1}$ &  \\
D & Hard & R & BM / P & 1.1879 & 0.0359 & 5.6922 & $<10^{-3}$ & *** \\
D & Hard & R & BM / E & 1.0129 & 0.0236 & 0.5513 & $5.8\times 10^{-1}$ &  \\
D & Hard & R & P / E & 0.8527 & 0.0276 & -4.9276 & $<10^{-3}$ & *** \\
ND & Easy & O & BM < P & 1.4076 & 0.0605 & 7.9585 & $<10^{-3}$ & *** \\
ND & Easy & O & P < E & 1.0750 & 0.0430 & 1.8099 & $3.5\times 10^{-2}$ & * \\
D & Easy & O & BM < P & 1.4351 & 0.0671 & 7.7268 & $<10^{-3}$ & *** \\
D & Easy & O & P < E & 1.0310 & 0.0503 & 0.6259 & $2.7\times 10^{-1}$ &  \\
ND & Hard & O & BM < P & 1.4695 & 0.0440 & 12.8425 & $<10^{-3}$ & *** \\
ND & Hard & O & P < E & 1.1574 & 0.0312 & 5.4161 & $<10^{-3}$ & *** \\
D & Hard & O & BM < P & 1.4989 & 0.0432 & 14.0286 & $<10^{-3}$ & *** \\
D & Hard & O & P < E & 1.0053 & 0.0316 & 0.1681 & $4.3\times 10^{-1}$ &  \\
\hline
\end{tabular*}
\caption{Estimated marginal means contrasts (four decimals). Acronyms: S=START, R=REWARD, O=OTHER, ND=No Delay, D=Delay, P=Pause, E=Execution.}
\label{tab:emmeans_contrasts}
\end{table}


\begin{table}[ht]
\centering
\begin{tabular*}{\textwidth}{@{\extracolsep{\fill}}lrrrrl}
\hline
Variables Predicting\\ logit(re-fixation proportion) & Estimate & Std. & z-value & p-value & \\
\hline\\
\multicolumn{6}{l}{\textit{Before Movement}} \\
\hline\\
Intercept                                & $-2.821$                  & $0.1182$                  & $-23.873$  & $<10^{-9}$      & *** \\
Difficulty (hard)                        & $0.4408$                  & $0.1015$                  & $4.342$    & $1.4\times10^{-5}$        & *** \\
 delay condition                        & $0.4629$                  & $0.1274$                  & $3.633$    & $2.8\times10^{-4}$        & *** \\
Difficulty $\times$  delay              & $-0.3878$                 & $0.1424$                  & $-2.723$   & $0.0065$                  & ** \\
Total fixations                          & $0.0431$                  & $0.00229$                 & $18.819$   & $<10^{-9}$       & *** \\
Group Var (participant)                  & $0.1938$                  & & & & \\
\hline\\
\multicolumn{6}{l}{\textit{Pause}} \\
\hline\\
Intercept                                & $-4.958$                  & $0.1365$                  & $-36.328$  & $<10^{-9}$      & *** \\
Difficulty (hard)                        & $0.4305$                  & $0.1770$                  & $2.432$    & $0.0150$                  & * \\
 delay condition                        & $0.0656$                  & $0.1941$                  & $0.3379$   & $0.7355$                  &  \\
Difficulty $\times$  delay              & $0.2467$                  & $0.2538$                  & $0.9722$   & $0.3309$                  &  \\
Total fixations                          & $0.1018$                  & $0.00340$                 & $29.902$   & $<10^{-9}$      & *** \\
Group Var (participant)                  & $0.0255$                  & & & & \\
\hline\\
\multicolumn{6}{l}{\textit{Execution}} \\
\hline\\
Intercept                                & $-3.605$                  & $0.1097$                  & $-32.849$  & $<10^{-9}$      & *** \\
Difficulty (hard)                        & $0.1912$                  & $0.0802$                  & $2.384$    & $0.0171$                  & * \\
 delay condition                        & $-0.2848$                 & $0.0764$                  & $-3.727$   & $1.9\times10^{-4}$        & *** \\
Difficulty $\times$  delay              & $0.1859$                  & $0.1080$                  & $1.722$    & $0.0851$                  & . \\
Total fixations                          & $0.0575$                  & $0.00193$                 & $29.733$   & $<10^{-9}$      & *** \\
Group Var (participant)                  & $0.1203$                  & & & & \\
\hline\\
\end{tabular*}
\caption{logit(re-fixation proportion). GLMM estimates of fixed effects. The p-value significance code: $*** < 10^{-3}$, $10^{-3} < ** < 10^{-2}$, $10^{-2} < * < 0.05$, $0.05 < . < 0.1$.}
\label{tab:logit(Repeat Ratio)}
\end{table}

\FloatBarrier
\subsection*{Reconstructing the upcoming path from visual fixation sequences}
\label{sec:reconstruction_path}

\begin{figure}
    \centering
    \includegraphics[width= \linewidth]{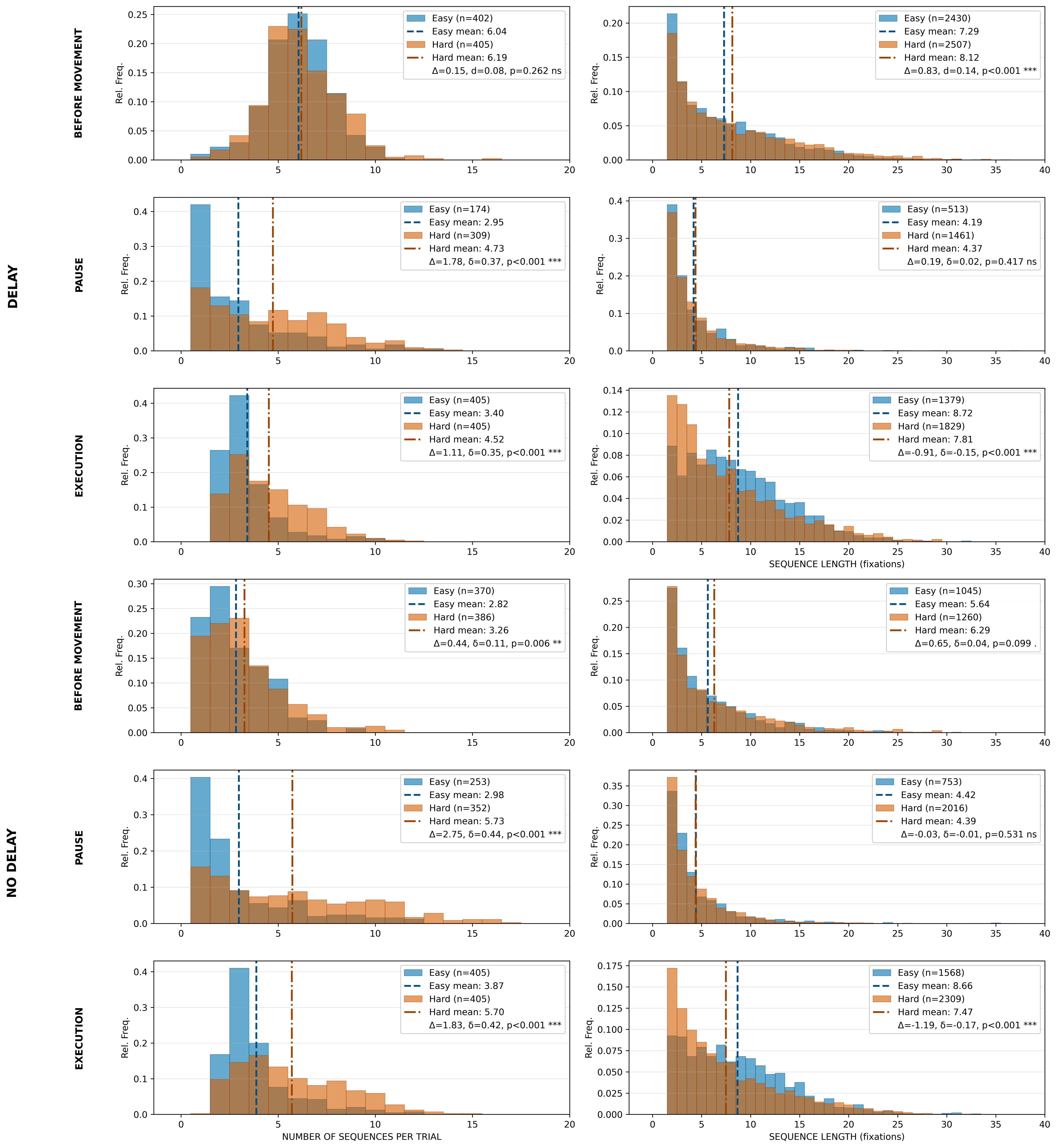}
    \caption{Distribution of the number of sequences and their lengths across all trials and participants. The distribution of (left column) the number and (right column) lengths of sequences identified by the clustering algorithm for each phase (Before Movement, Pause, Execution) for each  delay condition ( delay vs. No- delay), and for the two difficulties (Easy: blue, hard: orange). In each subplot the mean of the distribution is indicated by a different(ly styled) dashed line. A non parametric Mann--Whitney U test is performed on the mean of the distributions to test their differences. $\Delta$ is the difference between the means. The p-value significance code: $*** < 10^{-3}$,  $10^{-3} < ** < 10^{-2}$, $10^{-2} < * < 0.05$, $0.05 < . < 0.1$, $ > 0.1$. Effect sizes are reported with a Cohen's $d$ for the Before Movement/ delay condition; a Cliff's $\delta$ was used for all the other (strongly non-Gaussian) cases.}
    \label{fig:sequence_distribution}
\end{figure}

\begin{table}[ht]
\centering
\footnotesize
\begin{tabular*}{\textwidth}{@{\extracolsep{\fill}}lllrrrrl}
\hline
 delay Condition & Phase & Contrast & Odds Ratio & SE & z-ratio & p-value & sig \\
\hline\\[-2ex]
\multicolumn{8}{l}{\textbf{Fixation selection fraction} — fraction of fixations selected to be part of the best combination}\\[0.5ex]
NoDelay BM & Easy / Hard & 1.8438 & 0.0966 & 11.6784 & $<10^{-6}$ & *** \\
 Delay & BM & Easy / Hard & 1.3115 & 0.0327 & 10.8645 & $<10^{-6}$ & *** \\
NoDelay Pause & Easy / Hard & 1.9336 & 0.1002 & 12.7200 & $<10^{-6}$ & *** \\
 Delay & Pause & Easy / Hard & 1.7318 & 0.1064 & 8.9359 & $<10^{-6}$ & *** \\
NoDelay Exec & Easy / Hard & 3.0189 & 0.1228 & 27.1676 & $<10^{-6}$ & *** \\
 Delay & Exec & Easy / Hard & 2.9571 & 0.1580 & 20.2879 & $<10^{-6}$ & *** \\
\hline\\[-2ex]
\multicolumn{8}{l}{\textbf{No exclusion} — probability that no sequence is excluded (all sequences selected)}\\[0.5ex]
NoDelay BM & Easy / Hard & 2.0231 & 0.3124 & 4.5631 & $5.0\times 10^{-6}$ & *** \\
 Delay & BM & Easy / Hard & 1.0065 & 0.2494 & 0.0260 & $9.8\times 10^{-1}$ &  \\
NoDelay Pause & Easy / Hard & 3.6355 & 0.6411 & 7.3188 & $<10^{-6}$ & *** \\
 Delay & Pause & Easy / Hard & 3.0154 & 0.6040 & 5.5109 & $<10^{-6}$ & *** \\
NoDelay Exec & Easy / Hard & 4.5013 & 0.6854 & 9.8797 & $<10^{-6}$ & *** \\
 Delay & Exec & Easy / Hard & 4.2514 & 0.6958 & 8.8424 & $<10^{-6}$ & *** \\
\hline\\[-2ex]
\multicolumn{8}{l}{\textbf{Excluded on best} — fraction of excluded fixations falling on best-combination nodes}\\[0.5ex]
NoDelay BM & Easy / Hard & 0.6822 & 0.0758 & -3.4411 & $5.8\times 10^{-4}$ & *** \\
 Delay & BM & Easy / Hard & 1.0221 & 0.0584 & 0.3821 & $7.0\times 10^{-1}$ &  \\
NoDelay Pause & Easy / Hard & 0.9565 & 0.1075 & -0.3956 & $6.9\times 10^{-1}$ &  \\
 Delay & Pause & Easy / Hard & 0.8601 & 0.0993 & -1.3050 & $1.9\times 10^{-1}$ &  \\
NoDelay Exec & Easy / Hard & 0.7393 & 0.0823 & -2.7121 & $6.7\times 10^{-3}$ & ** \\
 Delay & Exec & Easy / Hard & 0.8410 & 0.1136 & -1.2820 & $2.0\times 10^{-1}$ &  \\
\hline\\[-2ex]
\multicolumn{8}{l}{\textbf{Direction distribution} — forward fraction (forward vs.\ reversed sequences)}\\[0.5ex]
NoDelay BM & Easy / Hard & 1.1331 & 0.1274 & 1.1109 & $2.7\times 10^{-1}$ &  \\
 Delay & BM & Easy / Hard & 1.4518 & 0.1256 & 4.3085 & $1.6\times 10^{-5}$ & *** \\
NoDelay Pause & Easy / Hard & 0.9094 & 0.1031 & -0.8375 & $4.0\times 10^{-1}$ &  \\
 Delay & Pause & Easy / Hard & 0.8329 & 0.1152 & -1.3218 & $1.9\times 10^{-1}$ &  \\
NoDelay Exec & Easy / Hard & 2.0173 & 0.3183 & 4.4473 & $8.7\times 10^{-6}$ & *** \\
 Delay & Exec & Easy / Hard & 2.1319 & 0.3539 & 4.5599 & $5.1\times 10^{-6}$ & *** \\
\hline
\end{tabular*}
\caption{Difficulty contrasts (Easy vs Hard) within each phase (BM, Pause, Exec) and
 delay condition, for the four dependent variables indicated by the section headers.
Odds ratios above 1 indicate larger values in easy than in hard maps.
Significance codes: *** $p<0.001$, ** $p<0.01$, * $p<0.05$.}
\label{tab:combined_difficulty_contrasts}
\end{table}

\FloatBarrier
\subsection*{Assessing the alignment between visual fixation sequences and the upcoming path: spatial coverage and temporal coherence}


\begin{table}[ht]
\centering
\footnotesize
\begin{tabular*}{\textwidth}{@{\extracolsep{\fill}}llrrrrl}
\hline
Held fixed & Comparison & $\Delta$ (logit) & SE & $z$ & $p$ & sig \\
\hline\\[-2ex]
\multicolumn{7}{l}{\textbf{Coverage} (adjusted cell means, difficulty averaged out)}\\[0.5ex]
Before mov. &  delay vs No- delay & $1.481$ & $0.037$ & $39.95$ & $<10^{-6}$ & *** \\
Execution &  delay vs No- delay & $-0.033$ & $0.036$ & $-0.92$ & $0.3574$ &  \\
 delay & BM vs Exec & $-0.557$ & $0.036$ & $-15.50$ & $<10^{-6}$ & *** \\
No  delay & BM vs Exec & $-2.071$ & $0.037$ & $-56.56$ & $<10^{-6}$ & *** \\
\hline\\[-2ex]
\multicolumn{7}{l}{\textbf{Coherence} (adjusted cell means, difficulty averaged out)}\\[0.5ex]
Before mov. &  delay vs No- delay & $0.025$ & $0.058$ & $0.44$ & $0.6603$ &  \\
Execution &  delay vs No- delay & $0.327$ & $0.052$ & $6.33$ & $<10^{-6}$ & *** \\
 delay & BM vs Exec & $-1.022$ & $0.051$ & $-19.92$ & $<10^{-6}$ & *** \\
No  delay & BM vs Exec & $-0.720$ & $0.057$ & $-12.65$ & $<10^{-6}$ & *** \\
\hline
\end{tabular*}
\caption{Contrasts between phase $\times$  delay-condition cells, within each measure. Each row either holds the phase fixed and compares  delay conditions, or holds the  delay condition fixed and compares phases. $\Delta$ is the difference between the covariate-adjusted cell means on the logit scale, with difficulty averaged out. Estimates come from a linear mixed model with a random intercept per participant and reference path length as covariate. $p$-value Holm-corrected within each measure. Significance codes: *** $p<0.001$, ** $p<0.01$, * $p<0.05$, . $p<0.1$.}
\label{tab:phase_waiting_contrasts}
\end{table}

\begin{table}[htbp]
\centering
\footnotesize
\begin{tabular}{ll l rrrr c r}
\toprule
Phase & DV & Path & Coef & SE & $z$ & $p$ & 95\% Boot CI & Prop.\ Med. \\
\midrule
  Before mov. & Coverage & $c$ (total) & 0.140 & 0.042 & 3.34 & $<$.001*** &  &  \\
   &  & $a$ (IV$\rightarrow$M) & 2.641 & 0.838 & 3.15 & 0.002** &  &  \\
   &  & $b$ (M$\rightarrow$DV) & 0.036 & 0.001 & 28.33 & $<$.001*** &  &  \\
   &  & $c'$ (direct) & 0.046 & 0.030 & 1.53 & 0.127 &  &  \\
   &  & $a \times b$ (indirect) & 0.094 & 0.030 & 3.13 & 0.002** & [0.019, 0.166] & 0.67 \\
\addlinespace[4pt]
\cmidrule(lr){3-9}
\addlinespace[2pt]
   & Coherence & $c$ (total) & 0.104 & 0.056 & 1.84 & 0.065$.$&  &  \\
   &  & $a$ (IV$\rightarrow$M) & 2.629 & 0.801 & 3.28 & 0.001** &  &  \\
   &  & $b$ (M$\rightarrow$DV) & -0.016 & 0.002 & -6.74 & $<$.001*** &  &  \\
   &  & $c'$ (direct) & 0.146 & 0.055 & 2.67 & 0.008** &  &  \\
   &  & $a \times b$ (indirect) & -0.043 & 0.015 & -2.95 & 0.003** & [-0.082, -0.011] & -0.41 \\
\addlinespace[6pt]
\cmidrule(lr){3-9}
\addlinespace[3pt]
  Execution & Coverage & $c$ (total) & 0.110 & 0.028 & 3.96 & $<$.001*** &  &  \\
   &  & $a$ (IV$\rightarrow$M) & 1.105 & 0.491 & 2.25 & 0.024* &  &  \\
   &  & $b$ (M$\rightarrow$DV) & 0.032 & 0.002 & 18.66 & $<$.001*** &  &  \\
   &  & $c'$ (direct) & 0.075 & 0.023 & 3.22 & 0.001** &  &  \\
   &  & $a \times b$ (indirect) & 0.035 & 0.016 & 2.23 & 0.026* & [0.003, 0.065] & 0.32 \\
\addlinespace[4pt]
\cmidrule(lr){3-9}
\addlinespace[2pt]
   & Coherence & $c$ (total) & 0.096 & 0.065 & 1.48 & 0.138 &  &  \\
   &  & $a$ (IV$\rightarrow$M) & 1.105 & 0.491 & 2.25 & 0.024* &  &  \\
   &  & $b$ (M$\rightarrow$DV) & -0.031 & 0.004 & -7.04 & $<$.001*** &  &  \\
   &  & $c'$ (direct) & 0.131 & 0.063 & 2.08 & 0.038* &  &  \\
   &  & $a \times b$ (indirect) & -0.035 & 0.016 & -2.14 & 0.032* & [-0.070, -0.007] & -0.36 \\
\addlinespace[6pt]
\cmidrule(lr){3-9}
\addlinespace[3pt]
  Pauses & Coverage & $c$ (total) & 0.500 & 0.131 & 3.80 & $<$.001*** &  &  \\
   &  & $a$ (IV$\rightarrow$M) & 3.549 & 0.874 & 4.06 & $<$.001*** &  &  \\
   &  & $b$ (M$\rightarrow$DV) & 0.093 & 0.005 & 17.95 & $<$.001*** &  &  \\
   &  & $c'$ (direct) & 0.178 & 0.103 & 1.74 & 0.082 $.$ &  &  \\
   &  & $a \times b$ (indirect) & 0.329 & 0.083 & 3.96 & $<$.001*** & [0.156, 0.513] & 0.66 \\
\addlinespace[4pt]
\cmidrule(lr){3-9}
\addlinespace[2pt]
   & Coherence & $c$ (total) & 0.661 & 0.191 & 3.46 & $<$.001*** &  &  \\
   &  & $a$ (IV$\rightarrow$M) & 3.499 & 0.979 & 3.57 & $<$.001*** &  &  \\
   &  & $b$ (M$\rightarrow$DV) & -0.029 & 0.010 & -2.90 & 0.004** &  &  \\
   &  & $c'$ (direct) & 0.761 & 0.192 & 3.95 & $<$.001*** &  &  \\
   &  & $a \times b$ (indirect) & -0.102 & 0.045 & -2.25 & 0.024* & [-0.218, -0.025] & -0.15 \\
\bottomrule
\end{tabular}
\vspace{4pt}
\begin{minipage}{\textwidth}
\scriptsize
\caption{Mediation analysis ( delay condition): Sequential mixed-model mediation analysis. The structure of the mediation analysis is Independent Variable (IV) $\rightarrow$ Mediator (M) $\rightarrow$  Dependent Variable (DV). IV is always the Difficulty; M is always the total number of fixations (for that phase); DV can be either coverage or coherence. All DVs are logit-transformed, as they are bound to be in the [0,1] interval. Random intercepts per participant are used in all paths LMM.
$c$ = total effect of difficulty; $a$ = effect of difficulty on mediator (total fixations); $b$ = effect of mediator on DV controlling for difficulty; $c'$ = direct effect of difficulty controlling for mediator; $a \times b$ = indirect effect. Sobel test used for indirect-effect $z$ and $p$. Bootstrap 95\% CI from participant-level resampling. Med.\ = proportion of total effect mediated ($a \times b / c$). The p-value significance (sig) code: $*** < 10^{-3}$, $10^{-3} < ** < 10^{-2}$, $10^{-2} < * < 0.05$, $0.05 < . < 0.1$.}
\label{tab:mediation_waiting}
\end{minipage}
\end{table}

\begin{table}[htbp]
\centering
\label{tab:mediation_no_waiting}
\footnotesize
\begin{tabular}{ll l rrrr c r}
\toprule
Phase & DV & Path & Coef & SE & $z$ & $p$ & 95\% Boot CI & Prop.\ Med. \\
\midrule
  Before mov. & Coverage & $c$ (total) & 0.295 & 0.075 & 3.92 & $<$.001*** &  &  \\
   &  & $a$ (IV$\rightarrow$M) & 3.732 & 0.868 & 4.30 & $<$.001*** &  &  \\
   &  & $b$ (M$\rightarrow$DV) & 0.062 & 0.002 & 26.76 & $<$.001*** &  &  \\
   &  & $c'$ (direct) & 0.070 & 0.056 & 1.25 & 0.210 &  &  \\
   &  & $a \times b$ (indirect) & 0.233 & 0.055 & 4.25 & $<$.001*** & [0.133, 0.329] & 0.79 \\
\addlinespace[4pt]
\cmidrule(lr){3-9}
\addlinespace[2pt]
   & Coherence & $c$ (total) & -0.039 & 0.145 & -0.27 & 0.786 &  &  \\
   &  & $a$ (IV$\rightarrow$M) & 2.999 & 1.064 & 2.82 & 0.005** &  &  \\
   &  & $b$ (M$\rightarrow$DV) & -0.008 & 0.005 & -1.52 & 0.127 &  &  \\
   &  & $c'$ (direct) & -0.013 & 0.146 & -0.09 & 0.931 &  &  \\
   &  & $a \times b$ (indirect) & -0.025 & 0.019 & -1.34 & 0.180 & [-0.081, 0.002] & 0.64 \\
\addlinespace[6pt]
\cmidrule(lr){3-9}
\addlinespace[3pt]
  Execution & Coverage & $c$ (total) & 0.112 & 0.030 & 3.73 & $<$.001*** &  &  \\
   &  & $a$ (IV$\rightarrow$M) & 0.342 & 0.528 & 0.65 & 0.518 &  &  \\
   &  & $b$ (M$\rightarrow$DV) & 0.030 & 0.002 & 16.95 & $<$.001*** &  &  \\
   &  & $c'$ (direct) & 0.103 & 0.026 & 3.93 & $<$.001*** &  &  \\
   &  & $a \times b$ (indirect) & 0.010 & 0.016 & 0.65 & 0.518 & [-0.014, 0.039] & 0.09 \\
\addlinespace[4pt]
\cmidrule(lr){3-9}
\addlinespace[2pt]
   & Coherence & $c$ (total) & 0.186 & 0.053 & 3.52 & $<$.001*** &  &  \\
   &  & $a$ (IV$\rightarrow$M) & 0.342 & 0.528 & 0.65 & 0.518 &  &  \\
   &  & $b$ (M$\rightarrow$DV) & -0.023 & 0.003 & -6.65 & $<$.001*** &  &  \\
   &  & $c'$ (direct) & 0.194 & 0.051 & 3.77 & $<$.001*** &  &  \\
   &  & $a \times b$ (indirect) & -0.008 & 0.012 & -0.64 & 0.520 & [-0.034, 0.014] & -0.04 \\
\addlinespace[6pt]
\cmidrule(lr){3-9}
\addlinespace[3pt]
  Pauses & Coverage & $c$ (total) & 0.182 & 0.093 & 1.97 & 0.049* &  &  \\
   &  & $a$ (IV$\rightarrow$M) & 2.909 & 0.892 & 3.26 & 0.001** &  &  \\
   &  & $b$ (M$\rightarrow$DV) & 0.065 & 0.003 & 20.75 & $<$.001*** &  &  \\
   &  & $c'$ (direct) & 0.001 & 0.072 & 0.01 & 0.989 &  &  \\
   &  & $a \times b$ (indirect) & 0.190 & 0.059 & 3.22 & 0.001** & [0.036, 0.328] & 1.04 \\
\addlinespace[4pt]
\cmidrule(lr){3-9}
\addlinespace[2pt]
   & Coherence & $c$ (total) & -0.018 & 0.162 & -0.11 & 0.913 &  &  \\
   &  & $a$ (IV$\rightarrow$M) & 3.890 & 1.068 & 3.64 & $<$.001*** &  &  \\
   &  & $b$ (M$\rightarrow$DV) & -0.016 & 0.007 & -2.31 & 0.021* &  &  \\
   &  & $c'$ (direct) & 0.035 & 0.161 & 0.22 & 0.827 &  &  \\
   &  & $a \times b$ (indirect) & -0.061 & 0.031 & -1.95 & 0.051$.$& [-0.128, -0.011] & 3.44 \\
\bottomrule
\end{tabular}
\vspace{4pt}
\begin{minipage}{\textwidth}
\caption{Mediation analysis (No- delay condition): Sequential mixed-model mediation analysis. The structure of the mediation analysis is Independent Variable (IV) $\rightarrow$ Mediator (M) $\rightarrow$  Dependent Variable (DV). IV is always the Difficulty; M is always the total number of fixations (for that phase); DV can be either coverage or coherence. All DVs are logit-transformed, as are bound to be in the [0,1] interval. Random intercepts per participant are used in all paths LMM.
$c$ = total effect of difficulty; $a$ = effect of difficulty on mediator (total fixations); $b$ = effect of mediator on DV controlling for difficulty; $c'$ = direct effect of difficulty controlling for mediator; $a \times b$ = indirect effect. Sobel test used for indirect-effect $z$ and $p$. Bootstrap 95\% CI from participant-level resampling. Med.\ = proportion of total effect mediated ($a \times b / c$). The p-value significance (sig) code: $*** < 10^{-3}$, $10^{-3} < ** < 10^{-2}$, $10^{-2} < * < 0.05$, $0.05 < . < 0.1$.}
\label{tab:mediation_no_waiting}
\end{minipage}
\end{table}

\clearpage

\subsection*{Sensitivity Analysis}
As described in the main Methods, sensitivity analyses were conducted by fixing the residual mediator--outcome correlation, $\rho_{\varepsilon}$, at successive values over a predefined grid and re-estimating the remaining model parameters by maximum likelihood. This procedure was applied to the significant indirect effects involving path coverage, coherence, and the number of backtracks. The critical value $\rho_{\varepsilon}^{*}$ was obtained by linearly interpolating between adjacent grid values at which the estimated mediator--outcome path changed sign. Different likelihood specifications were used for continuous and count outcomes. For coverage and coherence, both the mediator and the logit-transformed outcome were modelled as linear variables with correlated Gaussian residuals. Participant-level random intercepts were included in both equations and integrated out analytically, yielding an exact marginal likelihood that was maximized numerically. For the number of backtracks, the mediator was modelled using a linear Gaussian equation, whereas the outcome was modelled using a Poisson distribution with a log-linear mean. An additional Gaussian residual term was included in the outcome equation to accommodate overdispersion and to introduce the residual correlation $\rho_{\varepsilon}$ with the mediator equation. Because the resulting marginal likelihood had no closed-form solution, it was approximated through nested Gauss--Hermite quadrature, integrating over the participant-level random intercepts and the outcome-level residual. At each fixed value of $\rho_{\varepsilon}$, all remaining free parameters -- including the structural coefficients and variance components -- were re-estimated. 
The critical value for the backtrack analysis was $\rho_{\varepsilon}^{*} = -0.102$; \autoref{tab:critical_rho_eps_mediation_excl_p19} reports the corresponding values for coverage and coherence across task phases and  delay conditions.

\begin{table}[htbp]
\centering
\footnotesize
\begin{tabular*}{\textwidth}{@{\extracolsep{\fill}}lllr}
\toprule
Phase & DV &  delay condition & $\rho_{\varepsilon}^{*}$ \\
\midrule
Before mov. & Coverage & No  delay & 0.666 \\
 &  &  delay & 0.678 \\
\addlinespace[4pt]
\cmidrule(lr){3-4}
\addlinespace[2pt]
 & Coherence & No- delay &  ---  \\
 &  &  delay & -0.237 \\
\addlinespace[6pt]
\cmidrule(lr){3-4}
\addlinespace[3pt]
Execution & Coverage & No- delay & --- \\
 &  &  delay & 0.539 \\
\addlinespace[4pt]
\cmidrule(lr){3-4}
\addlinespace[2pt]
 & Coherence & No. delay & --- \\
 &  &  delay & -0.248 \\
\addlinespace[6pt]
\cmidrule(lr){3-4}
\addlinespace[3pt]
Pauses & Coverage & No- delay & 0.626 \\
 &  &  delay & 0.620 \\
\addlinespace[4pt]
\cmidrule(lr){3-4}
\addlinespace[2pt]
 & Coherence & No- delay & --- \\
 &  &  delay & -0.144 \\
\bottomrule
\end{tabular*}
\vspace{4pt}
\begin{minipage}{\textwidth}
\caption{Sensitivity analysis (mediation cells): critical value $\rho_{\varepsilon}^{*}$ of the residual mediator--outcome correlation at which the mediation path $b$ changes sign, by phase, dependent variable (DV), and  delay condition (participant P19 excluded). IV is always Difficulty; M is always the total number of fixations (for that phase); DV is Coverage or Coherence (logit-transformed). $\rho_{\varepsilon}^{*}$ is obtained by sweeping the assumed residual correlation and linearly interpolating the value at which $b$ crosses zero.}
\label{tab:critical_rho_eps_mediation_excl_p19}
\end{minipage}
\end{table}

\end{document}